\documentclass[a4paper,11pt,british]{article}
\usepackage{jheppub_bibtex} 

\usepackage{url}            
\usepackage{booktabs}       
\usepackage{amsfonts}       
\usepackage{nicefrac}       
\usepackage{microtype}      
\usepackage{xcolor}         
\usepackage{amsmath}

\usepackage{tabularx}
\usepackage{wrapfig}
\usepackage{amssymb}
\usepackage{amsthm}
\usepackage{multicol}
\usepackage{multirow}
\usepackage{csquotes}
\usepackage{siunitx}
\usepackage{enumitem}
\usepackage[capitalise]{cleveref}  
\usepackage[symbol]{footmisc}
\usepackage{tablefootnote}
\usepackage{accents}  
\usepackage{mathtools}  
\usepackage{pgfplots, pgfplotstable}
\usepgfplotslibrary{groupplots, fillbetween}

\usepackage{tikz}
\usetikzlibrary{arrows.meta,
                calc, 
                positioning,
                decorations.pathmorphing,
                decorations.pathreplacing,
                tikzmark,
                shapes.multipart,
                arrows,
                shadows,
                angles,
                quotes,
                patterns}

\usepackage{multirow}

\usepackage[
  backend=biber,
  style=numeric-comp,
  sorting=none,
  sortcites=true,
  maxcitenames=1,
  maxbibnames=999,
  url=false,
  doi=true,
  isbn=false,
  eprint=false,
  giveninits=true 
]{biblatex}

\AtEveryBibitem{%
  \clearfield{note}%
}

\newcommand{\fl}[0]{_\flat} 
\newcommand{\FS}[0]{_\mathrm{FS}}
\newcommand{\dd}[0]{\partial\overline{\partial}}

\newcommand{\ZZ}[0]{\mathbb{Z}}

\newcommand{\RR}[0]{\mathbb{R}}
\newcommand{\CC}[0]{\mathbb{C}}

\newcommand{\PP}[0]{\mathbb{P}}
\renewcommand{\SS}[0]{\mathbb{S}}

\DeclarePairedDelimiter{\paren}{\lparen}{\rparen}
\DeclarePairedDelimiter{\abs}{\lvert}{\rvert}
\DeclarePairedDelimiter{\norm}{\lVert}{\rVert}

\DeclareMathOperator{\tr}{tr}

\DeclareMathOperator{\Sym}{Sym}
\DeclareMathOperator{\Isom}{Isom}
\DeclareMathOperator{\diag}{diag}

\newtheorem*{theorem*}{Theorem}
\newtheorem{theorem}{Theorem}[section]

\newtheorem{assumption}{Assumption}[section]
\newtheorem{lemma}[theorem]{Lemma}
\newtheorem{prop}[theorem]{Proposition}
\newtheorem{conj}[theorem]{Conjecture}

\crefname{prop}{Proposition}{Propositions}

\newcommand{\deepBlue}{\color[HTML]{5975A4}}
\newcommand{\deepOrange}{\color[HTML]{CC8963}}
\newcommand{\deepGreen}{\color[HTML]{5f9E6E}}
\newcommand{\deepRed}{\color[HTML]{B55D60}}
\definecolor{deepBlue}{HTML}{5975A4}
\definecolor{deepOrange}{HTML}{CC8963}
\definecolor{deepGreen}{HTML}{5f9E6E}
\definecolor{deepRed}{HTML}{B55D60}

\definecolor{deepPurple}{HTML}{8172B3}

\title{\boldmath Symbolic Approximations to Ricci-flat Metrics Via Extrinsic Symmetries of Calabi--Yau Hypersurfaces}

\author{Viktor Mirjani\'c}
\author{and Challenger Mishra}
\affiliation{Department of Computer Science and Technology, University of Cambridge}
\emailAdd{vvm22@cam.ac.uk}
\emailAdd{cm2099@cam.ac.uk}

\abstract{%
Ever since Yau's non-constructive existence proof of Ricci-flat metrics on Calabi--Yau manifolds, finding their explicit construction remains a major obstacle to development of both string theory and algebraic geometry. 
Recent computational approaches employ machine learning to create novel neural representations for approximating these metrics, offering high accuracy but limited interpretability.
In this paper, we analyse machine learning approximations to flat metrics of Fermat Calabi--Yau $n$-folds and some of their one-parameter deformations in three dimensions in order to discover their new properties. We formalise cases in which the flat metric has more symmetries than the underlying manifold, and prove that these symmetries imply that the flat metric admits a surprisingly compact representation for certain choices of complex structure moduli. We show that such symmetries uniquely determine the flat metric on certain loci, for which we present an analytic form. We also incorporate our theoretical results into neural networks to reduce Ricci curvature for multiple Calabi--Yau manifolds compared to previous machine learning approaches. We conclude with distilling the ML models to obtain for the first time closed form expressions for K\"ahler metrics with near-zero scalar curvature.}

\pgfplotsset{compat=1.18}
\makeatletter
\gdef\@fpheader{}
\makeatother
\begin{document}
\maketitle
\flushbottom

\section{Introduction}

Since the last century, algebraic geometry has been closely tied to modern physics via string theory in an attempt to unify gravity and quantum mechanics. One of the major open problems, relevant to both algebraic geometry and superstring theory, involves the study of Ricci-flat metrics on Calabi--Yau varieties. The seminal Calabi--Yau theorem \cite{yau_ricci_1978} states that these manifolds possess a unique flat metric in each K\"ahler class. This metric is essential for computing phenomenological quantities, such as particle masses and physical Yukawa couplings in the context of Calabi--Yau compactifications of the heterotic string, but an explicit construction for it has not been found yet.

In this paper we do not focus on these downstream uses of Ricci-flat metrics. Instead, we make finding them less challenging, as very little is known about them in the first place. There exists an algorithm to compute them due to \textcite{donaldson_scalar_2001, donaldson_numerical_2005}, but it suffers from the curse of dimensionality. A refinement of this approach by \textcite{headrick_energy_2010} is still state-of-the-art, but does not scale beyond a small set of highly-symmetric manifolds. Neural networks achieve good results more generally, but have other issues such as lack of convergence guarantees (see \cref{sec:ml_cy} and \cite{ek_calabi-yau_2024} for a more detailed discussion). We show that the neural network outputs can be analysed to yield new properties of the flat metric and its symmetries.

Studying symmetries of families of Calabi--Yaus has long been a fruitful endeavour yielding many interesting results. One remarkable observation with many far-reaching consequences for string theory and for algebraic geometry is the mirror symmetry of \textcite{candelas_pair_1991} that relates Calabi--Yaus with their mirror pairs. On the other hand, one can study symmetries of a single manifold via its isometry group. These symmetry groups have been extensively studied \cite{braun_calabi-yau_2008, braun_three-generation_2010, davies_calabi-yau_2010, braun_free_2011, mishra_calabi-yau_2017, candelas_calabi-yau_2018, lukas_discrete_2020}, often in the context of building string vacua that recover the standard model of particle physics. In three complex dimensions, such symmetries are discrete and can improve understanding of a myriad phenomenological issues that arise when connecting string theory to four-dimensional particle physics \cite{anderson_heterotic_2008}, such as proton decay (by studying R symmetries) and Wilson line breaking for gauge unification. They can also help in explaining the structure of flavour mixing matrices such as CKM and PMNS \cite{constantin_fermion_2024} and in studying anomaly cancellation \cite{ibanez1992discrete}. Furthermore, one can quotient the manifolds by free group actions to obtain new Calabi--Yaus with different topological properties for string compactification \cite{candelas_new_2010, candelas_highly_2018, candelas2016hodge,candelas_calabi-yau_2018,constantin2016family}.

Our contributions are that we look at extensions of $(0,0)$-forms on Calabi--Yau hypersurfaces to the ambient space $\PP^n$ in order to describe cases in which flat metrics have representations with additional symmetries that were not previously considered.

These symmetries are separate from the symmetries of the underlying Calabi--Yaus, and they can be continuous even if the Calabi--Yau only has discrete symmetries. There are no theoretical obstructions to this since these symmetries exist only in the ambient space. Thus, for example, the vanishing of the Killing vector field for some Calabi--Yaus only implies non-existence of continuous symmetries on the Calabi--Yau itself, but says nothing about these \emph{extrinsic} symmetries that we introduce with our paper.

We show that these additional symmetries strongly constrain the flat metric. This results in easier training of neural networks, and improves upon similar observations first made by \textcite{jejjala_neural_2021}. It also allows for distillation into compact closed form expressions without increase in loss. One major issue with ML approximated flat metrics is that certain constraints, such as positivity, are often only softly enforced. By distilling ML models into closed form we circumvent many downsides such as this, allowing them to be directly checked if needed. Finally, the symmetry conditions we identify are strong enough that there emerge non-trivial loci on the Calabi--Yaus, on which the flat metric becomes completely determined.

The outline of the paper is as follows. In \cref{sec:background} we will summarise machine learning approaches for learning Ricci-flat metrics. For a more detailed discussion, we refer the reader to \textcite[Sections 2 and 3]{ek_calabi-yau_2024}. In \cref{sec:encoding_symmetries} we motivate our main results by showing that symmetry group size is related to salience in learnt NN approximations, and discuss how various symmetries of Calabi--Yau metrics can be incorporated into ML models with Geometric Deep Learning formalism of \textcite{bronstein_geometric_2021}. In \cref{sec:hidden_sym} we state our main theoretical results, namely a phenomenon that we call the \emph{extrinsic symmetries} of Calabi--Yau manifolds, whereby both the K\"ahler potential of the flat metric and its related differential forms exhibit representations with strictly more symmetries than the manifold itself.\footnote{The name is thus a slight misnomer as \textquote{extrinsic} refers not to the manifold but to differential forms.} In \cref{sec:phi_conjecture} we experimentally validate our hypothesis on multiple families of Calabi--Yaus. We show that our approach achieves lower loss with much less compute power on the Fermat Quintic compared to previous machine learning approaches. Subsequently, we focus on extracting symbolic expressions for corrections to the K\"ahler potential from the machine derived models. In \cref{sec:phi_symb} we describe closed-form symbolic approximations that also exhibit low loss. We close off with \cref{sec:exact_formulae} where we address the challenge of finding \emph{exact formulae} for the Ricci-flat metric, and show that there exist non-trivial loci on the Calabi--Yaus where the flat metric is completely determined by its symmetries.

\section{Background and Related Work}
\label{sec:background}
\subsection{Machine Learning}

Neural networks are \emph{universal approximators}, meaning that in the limit of infinite size they can approximate any reasonably well-behaved function to an arbitrary degree of accuracy~\cite{cybenko_approximation_1989, hornik_multilayer_1989}. For example, a two-layer MLP (Multi-Layer Perceptron) with non-polynomial activation is an universal approximator for any continuous, bounded function on a compact subset of $\RR^n$ \cite{hornik_approximation_1991}. On the other hand, a Graph Neural Network (GNN) such as GIN \cite{  xu_how_2019} is a universal approximator for similarly restricted functions that operate on graph inputs. There are a plethora of universal approximation guarantees for various other neural architectures. However, these guarantees may not reveal much about the \emph{training process} itself or determining the optimal network parameters.

Due to their powerful approximating capabilities, neural networks can be used to solve regression tasks even when the ground truth is unknown in an explicit sense. When learning a function $f(x)=y$ using a dataset of $(x,y)$ pairs, we can set up a loss function like $\mathcal{L}=\abs{\hat y - y}$ or similar, where we use the caret $\hat{\paren{\cdot}}$ to distinguish a neural network approximation $\hat f(x)=\hat y$ from the true function. If $y$ is unknown, but we know that $f$ satisfies a condition $C(f(x))=0$, then we can adapt the loss to $\mathcal{L}=\abs{C(\hat y)}$. This is the guiding principle of physics inspired neural networks (PINNs) \cite{raissi_physics_2017, raissi_physics-informed_2019}.

PINNs approximately solve PDEs using a loss function that measures how well the neural network satisfies differential equations and initial conditions. Depending on the nature of the PDE and complexity of the underlying space, training can be a challenging exercise. 

\subsection{Calabi--Yau Manifolds}

We define a Calabi--Yau manifold, or $n$-fold, to be a compact K\"ahler manifold of complex dimension $n$ that has a nowhere vanishing holomorphic $n$-form. This form is known as the volume or top form and is denoted by $\Omega$. Its existence implies that the first Chern class $c_1$ of the Calabi--Yau manifold vanishes. A seminal result by \textcite{yau_ricci_1978} showed that these manifolds admit a unique metric with vanishing Ricci tensor in each of its K\"ahler classes. 

Calabi--Yau manifolds are a cornerstone of superstring theory. In the 1980s, it was realised that these geometries play an important role in compactifying a higher dimensional theory of quantum gravity to reveal details of the four-dimensional physical universe \cite{candelas_vacuum_1985}. Although there are a number of challenges in recovering the details of our physical universe through string compactifications, understanding Calabi--Yau manifolds is a key step towards making increasingly accurate phenomenological predictions in heterotic string theory. 

In some settings, broader features of the particle spectra depend only on the \emph{topology} of the Calabi--Yau geometry. For example, number of generations of particles one recovers in the standard embedding is $|\chi|/2$, where $\chi$ is the Euler characteristic of the manifold --- a topological feature. Therefore, only some compactification settings will result in three generations of matter particles that we observe in the Standard Model of particle physics. However, the choices are many, and while the exact number of Calabi--Yau (CY) manifolds is unknown, the number of compactification settings can be as large as $10^{\num{500}}$ \cite{denef_distributions_2004}. On the other hand, many other features of the four-dimensional particle spectra such as masses of fermions or physical coupling constants require not just topological features, but also the Ricci-flat metric of the underlying Calabi--Yau geometry. Therefore, a better understanding of such metrics is essential for progress in this field. 

There are many ways to obtain a Calabi--Yau manifold, and a hypersurface construction is one of the most straightforward. A \emph{hypersurface} is a zero locus of any (homogeneous) polynomial $Q$ in $\PP^n$, but it is Calabi--Yau only if $\deg Q=n+1$, with dimension $d=n-1$. This can be further generalised into complete intersection Calabi--Yaus (CICYs). Complete intersection varieties are a joint zero locus of multiple polynomials in an ambient space made of multiple projective spaces. The Calabi--Yau condition for complete intersections is $\sum_j \deg_i Q_j = n_i+1$, where the equation holds for each projective space in the ambient space $\PP^{n_1}\times\PP^{n_2}\times\dots\times\PP^{n_k}$, and $\deg_i Q_j$ is the degree of the $j$-th polynomial with respect to the $i$-th projective space \cite{candelas_de_la_ossa, hubsch_calabi-yau_1994}. One can easily see that the Calabi--Yau condition for hypersurfaces is a special case of that for CICYs.

In this paper we focus exclusively on Calabi--Yau hypersurfaces, but we expect that our methods can be extended to CICYs as well. Specifically, we focus on the Fermat family of Calabi--Yaus, defined as the zero locus of the following polynomial
\begin{equation}
\label{eq:fermat_general_homo}
    Q_n\!:\ Z_0^{n+1}+Z_1^{n+1}+\dots+Z_n^{n+1}=0 
\end{equation}
One often normalises this equation with $Z_0=1$ and $z_i=Z_i/Z_0$ obtaining
\begin{equation}
\label{eq:fermat_general}
    Q_n\!:\ 1+z_1^{n+1}+z_2^{n+1}+\dots+z_n^{n+1}=0 
\end{equation}
In ambient spaces $\PP^n$ for $n=2,3,4$, we define the Complex Torus, K3 manifold, and the Fermat Quintic as 
\begin{subequations}
\begin{align}
    Q_\mathrm{Torus}\!:&\ 1+z_1^3+z_2^3 = 0 \label{eq:torus} \\
    Q_\mathrm{K3}\!:&\ 1+z_1^4+z_2^4+z_3^4 = 0 \label{eq:k3} \\
    Q_\mathrm{Quintic}\!:&\ 1+z_1^5+z_2^5+z_3^5+z_4^5 = 0 \label{eq:quintic}
\end{align}
\end{subequations}
which have dimension $d=1,2,3$, respectively. One can keep increasing the dimension, obtaining a  sextic $4$-fold, and so on, and the relation $n=d+1$ always gets preserved. The simplicity and symmetry of these polynomials has made Fermat $d$-folds a good benchmark for experiments \cite{jejjala_neural_2021, douglas_numerical_2008, douglas_numerical_2007, larfors_learning_2021, hendi_learning_2024, berglund_machine_2023, ek_calabi-yau_2024, ashmore_machine_2020, gerdes_cyjax_2022, headrick_energy_2010, headrick_numerical_2005, anderson_lectures_2023, butbaia_physical_2024, witten_symmetry_1985, anderson_moduli-dependent_2021}, even though phenomenologically they might not be an exact match for the observable universe.

\subsection{Machine Learning Ricci-flat Metrics}
\label{sec:ml_cy}

On a Calabi--Yau one can define a \emph{metric} $g$, a matrix-valued function that satisfies conditions such as K\"ahlerity ($\partial_k g_{ij}=\partial_i g_{kj}$) and positivity ($g>0$). To be exact, a K\"ahler metric also needs to be positive in addition to satisfying the derivative condition, which is a subtle but important point. To give an example, one can recall the Fubini-Study metric $g\FS=I/\norm{Z}^2 - Z^\dag \otimes Z / \norm{Z}^4$ on a projective space. Then, its pullback $\iota^* g\FS$ is a K\"ahler metric on the Calabi--Yau.

We are motivated by physics to add another condition to the metric, Ricci-flatness, and obtain a metric that has vanishing curvature. This condition simplifies to $\dd \log\det g = \mathbf{0}_d$ for K\"ahler metrics. It is known from theory that the Ricci-flat metrics are rare among the K\"ahler metrics. In the case of Fermat Calabi--Yaus, the flat metric is unique up to a constant scaling factor.

Unfortunately, the original existence proof by Yau is non-constructive \cite{yau_ricci_1978} and for the longest time the flat metric was only known in the special case of toric varieties. The most notable recent progress is by \textcite{kachru_k3_2020} who produced the flat metric on K3, a non-toric Calabi--Yau. They gave not a closed-form expression but a perturbative method, and they only computed the metric to the leading order in FI parameter $\xi$, but they argued that computing higher orders is also possible and requires solving systems of linear equations only. Beyond tori and K3, no known methods to find the flat metric exist yet.

Fortunately, we can find good enough approximations by optimizing neural networks with gradient descent. However, to make this approach viable, several theoretical results are needed. Firstly, there exists a flat metric $g\fl$ in the same cohomology class as the pullbacked Fubini-Study metric $\iota^* g\FS$, meaning that we can write
\begin{equation}
    \label{eq:fs_corr}
    g\fl=\iota^*\left(g\FS+\dd\phi\right)
\end{equation}
where $\phi:\PP^n\to\RR$ is unique up to factors that vanish after $\iota^*\dd$, and $\iota^*$ is the pullback via Jacobians from homogeneous to local coordinates. Since $\phi$ is scalar-valued, optimizing it is much easier than $g$. This approach, named PhiModel by \textcite{larfors_learning_2021}, is the current standard to learn $g\fl$. Its another advantage is that it enforces the partial derivative constraint for K\"ahlerity on the neural network output by construction. A spectral network by \textcite{berglund_machine_2023} forces $\phi$ to be homogeneous by carefully adapting the neural architecture. We will revisit spectral models in the next section.

Second modification to make ML viable involves simplifying the loss function. A metric is Ricci-flat if its curvature vanishes, and a na\"ive approach would be to define the loss function $\mathcal{L}_{Ric}$ to be its absolute value. However, that would require taking fourth-order mixed derivatives of $\phi$, which makes it hard to backpropagate through.

Instead, a surrogate loss is employed. From theory, the flat metric satisfies the Monge--Amp\`ere equation:
\begin{equation}
    \label{eq:ma}
    \det g\fl = \kappa \Omega \overline{\Omega}
\end{equation}
where $\Omega$ is the volume form, and $\kappa$ is a real constant. Furthermore, $\Omega$ is easily computable using the defining polynomial, and $\kappa$ can be numerically approximated by integrating \cref{eq:ma} over the Calabi--Yau. Crucially, the implication is bidirectional: a K\"ahler metric satisfying the Monge--Amp\`ere equation is Ricci-flat, so we replace the Ricci loss $\mathcal{L}_{Ric}$ with the $\sigma$-measure
\begin{equation}
    \label{eq:sigma_measure}
    \mathcal{L}_\sigma = \abs*{1 - \frac{\det \hat g}{\det g\fl}} = \abs*{1 - \frac{\det \hat g}{ \kappa \Omega \overline{\Omega}}}
\end{equation}
Thus, the problem becomes finding $\phi$ that minimises $\mathcal{L}_\sigma$. This setup is similar to PINNs, which is to be expected since the fundamental problem of solving a PDE is shared between PINNs and this model. But, there are also notable differences, such as lack of initial conditions for the Ricci-flat metric, presence of additional structure and symmetries originating from the manifolds, and use of surrogate losses.

Using neural networks to approximate flat metrics, while well established by \cite{jejjala_neural_2021, ashmore_machine_2020, berglund_cymyc_2024, berglund_machine_2023, larfors_learning_2021, hendi_learning_2024} and others, nevertheless has several issues that are identified by \textcite{ek_calabi-yau_2024}. Firstly, the metric $g$ is also required to be positive definite, which is not directly enforced by generic neural networks. This would require the correction $\phi$ in the PhiModel to be \emph{pluri-subharmonic} \cite{berglund_machine_2023}, which is a non-trivial property to guarantee. Instead, the networks are biased to positive definiteness through careful initialisation: $g\FS>0$, so if we set $\phi_{init}\approx0$, and if the NN stays near its initialisation during training, then positive definiteness is likely conserved. However, there are no theoretical guarantees that this will happen in practice, and checking positivity \emph{post-hoc} is difficult since the metric can violate positivity on a set of measure zero. This is unsettling because much of the theoretical apparatus relies precisely on positivity assumptions.

The second issue is that the $\sigma$-measure is not a perfect surrogate for Ricci loss. Having $\mathcal{L}_\sigma$ be identically zero does imply that the metric is Ricci-flat, but there are no known convergence guarantees that minimizing one also reduces the other in the approximate case. In practice, there are instances of Calabi--Yaus with pathological metrics that have low $\mathcal{L}_\sigma$ but diverging curvature, as can be seen in e.g.\ \textcite{hendi_learning_2024} with their discontinuous approximations.

Despite these issues, neural networks are converging to low $\sigma$-loss orders of magnitude faster compared to other, more principled approaches such as the Donaldson's algorithm \cite{donaldson_numerical_2005} and its variants, and so they remain important when studying numerical flat metrics.

\subsection{Explainable AI}

Neural networks for all intents and purposes behave as black boxes, and their enormous parameter space makes it difficult to interpret how they produce their outputs. Explainable AI is a research area that tries to make NNs interpretable either by constructing interpretable architectures, or with \emph{post-hoc} techniques that can be applied to already trained models. In order to construct a model that is interpretable by design, it is useful to take knowledge of symmetries into account. 

Alternatively, one can completely abandon neural networks in favor of different, more explainable architectures. For example, Kolmogorov--Arnold Networks by \textcite{liu_kan_2024} replace linear weights and non-linear activation functions with learnable splines, with the effect that very small KANs can match the performance of MLPs with orders of magnitude more parameters. They also have universality guarantees from the Kolmogorov--Arnold representation theorem \cite{kolmogorov:superposition, arnold_representation_2009} similar to that of MLPs.

Yet another way to obtain interpretable formulae by construction is via symbolic regression. The best symbolic regression framework known to authors is PySR \cite{cranmerDiscovering2020, cranmerInterpretableMachineLearning2023}, which implements a genetic algorithm in Julia backend that searches for a symbolic function by combining supplied primitives such as addition, multiplication, exponentiation, etc. 

Meanwhile, the \emph{post-hoc} methods generally use some gradient information to give each feature an attribution score measuring its effect on the loss. Such attribution score, or \emph{salience}, in its simplest form is
\begin{equation}
\label{eq:salience}
    a_i = \frac{1}{n} \sum_{k=1}^n \abs*{\frac{\partial \mathcal{L}_\sigma^{(k)}}{\partial x_{i}^{(k)}}}
\end{equation}
where $i$ ranges over the features $x$, and $k$ over entries in the dataset.

There are many different ways to assign attribution scores to neural network inputs. One popular method is integrated gradients by \textcite{sundararajan_axiomatic_2017}. However, methods like this are more suited towards \textquote{real-world data}. Integrated gradients in particular require computation with respect to a chosen baseline. This works fine with e.g.\ images, where the authors recommend taking a black image or white noise as baseline, but it becomes less clear how to translate this to the synthetic dataset of Calabi--Yaus. Here inputs are in $\PP^n$, which does not have a mathematically distinguished origin point, so any choice of baseline would have to break the symmetry of the manifold in some way. Furthermore, PhiModel optimises for higher-order derivatives, so we can expect that simple gradient analysis already carries sufficient information about how inputs are used. And, since we use small models in our experiments, we do not suffer from vanishing gradients that might affect our attribution scores. Thus, the issues that necessitated creation of advanced methods like integrated gradients or LRP \cite{montavon_layer-wise_2019} are likely not applicable in this case. In fact, this simple gradient attribution of \cref{eq:salience} has already led to new discoveries in knot and representation theory in \textcite{davies_advancing_2021}.

\section{Symmetric Forms on Calabi--Yaus}
\label{sec:encoding_symmetries}

Calabi--Yau manifolds and their metrics are often highly symmetric, and a neural network that captures these symmetries will train more easily and achieve lower loss. Specifically, in this paper we look at Fermat Calabi--Yaus which have the following symmetries:

\begin{itemize}
    \item $\CC^*$ projective symmetry inherited from embedding into a $\PP^n$ ambient space,
    \item $\ZZ_2$ conjugation symmetry as $Q$ is invariant under conjugation,
    \item $\SS_n$ permutation symmetry as $Q$ is symmetric,
    \item $\ZZ_{n+1}^n$ toric symmetry as $Q$ is invariant under actions $Z_j \mapsto e^{\frac{2i\pi}{n+1} k_j} Z_j$, for $k_j\in \ZZ_{n+1}$.
\end{itemize}

\textcite{berglund_machine_2023} designed \textquote{spectral networks} that take matrix $A_{ij}=\left(\frac{Z_i\overline{Z_j}}{\norm{Z}^2}\right)$ as input. Owing to well known results in invariant theory \cite{villar_scalars_2023, weyl_classical_1946}, this network is invariant under the $\CC^*$ projective symmetry by construction, while also being able to approximate any $\CC^*$-symmetric function in the infinite limit.

In \textcite{hendi_learning_2024} more symmetries were added by arbitrarily defining one group element as fundamental, and mapping points on the CY to corresponding fundamental domains. This may create discontinuities at domain boundaries, and such networks may only be reducing $\sigma$-measure and not the underlying Ricci loss, but this may become addressed in the future by tweaking the architecture.  

We take a different approach and give a blueprint for how these symmetries can be treated \emph{naturally}, and in the spirit of Geometric Deep Learning of \textcite{bronstein_geometric_2021}. As mentioned above, the scaling symmetry is handled by using spectral networks. To capture invariance under toric symmetries it is sufficient to substitute coordinates $Z$ with invariant features $Z^n$, which immediately follows from the following statement.
\begin{prop}
\label{prop:f_rot_sym}
    Let $\zeta^n=1$ and $f:\CC\to\CC$ be a function with a rotation symmetry $f(z)=f(\zeta z)$. Then there exists $g:\CC\to\CC$ satisfying $f(z)=g(z^n)$.
    \begin{proof}
        Let $\pi:z\mapsto \abs{z}^{1/n}e^{i/n \arg z}$ be the principal $n$-th root. It satisfies $\pi(z^n)=z \zeta^k$ for some $k\in\ZZ$ as a function of $z$. Then, just take $g=f\circ \pi$, which is well defined and single--valued, and trivially satisfies $g(z^n)=f(\pi(z^n))=f(z \zeta^k)=f(z)$.
    \end{proof}
\end{prop}
Crucially, unlike the construction based on fundamental domains, the domain of $g$ is the entire complex plane. Therefore, enforcing continuity of $f$ is easier here, whereas it would introduce boundary conditions if one used fundamental domains. To capture conjugation invariance, we can similarly replace ($\Re$, $\Im$) maps in the spectral network either with ($\abs{\cdot}$,$\Re$) or with ($\Re$, $\abs{\Im}$), both of which respect conjugation symmetry. Thus, we are only left with permutation symmetry to handle.

\subsection{Permutation Symmetry of Calabi--Yaus}

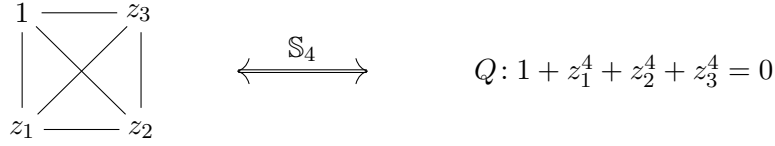
\begin{figure}[ht]
    \centering
    \begin{tikzpicture}
    \tikzstyle{vertex}=[circle, inner sep = 1.5pt, minimum size = 5mm]
    
    \node[vertex]      (z1)                                           {$z_1$};
    \node[vertex]      (z2)       [right= of z1]                      {$z_2$};
    \node[vertex]      (one)      [above= of z1]                      {$1$};
    \node[vertex]      (zd)       [right= of one]                     {$z_3$};
    \node[vertex]      (zd_dummy) [above= of z2]                      {};
    \node              (mid)      [right= 2 of $(z2)!0.5!(zd_dummy)$] {};
    \node              (p)        [right= 2 of mid]                   {$ Q\!: 1+z_1^4+z_2^4+z_3^4=0$};
    
    \draw[-] (z1) -- (z2);
    \draw[-] (z1) -- (zd_dummy);
    \draw[-] (z1) -- (one);
    \draw[-] (z2) -- (zd_dummy);
    \draw[-] (z2) -- (one);
    \draw[-] (zd) -- (one);
    \draw[double,<->] ($(mid.east) + (-1,0)$) -- ($(mid.west) + (+1,0)$) node[midway,above] () {$\SS_4$};
\end{tikzpicture}   
    \caption{The symmetry group of a K3 defining polynomial (right), $\SS_4$, is also the automorphism group of the complete graph (left).}
    \label{fig:k3_graph_sym}
\end{figure}

\begin{figure}[ht]
    \centering
    \begin{tikzpicture}
    \tikzstyle{vertex}=[circle, inner sep = 1.5pt, minimum size = 5mm]

    \node[vertex]      (z1)                             {$z_1$};
    \node[vertex]      (z2)       [right=of z1]         {$z_2$};
    \node[vertex]      (z4)       [above= 0.5 of $(z1)!0.5!(z2)$]   {$z_4$};
    \node[vertex]      (z3)       [right=of z4]   {$z_3$};
    \node[vertex]      (w1)       [below=of z1]   {$w_1$};
    \node[vertex]      (w2)       [below=of z2]   {$w_2$};
    \node[vertex]      (w3)       [below=of z3]   {$w_3$};
    \node[vertex]      (w4)       [below=of z4]   {$w_4$};
    \node              (mid)      [right= 2 of w3, yshift=4mm] {};
    \node              (p)        [right= 2 of mid] {$\begin{aligned}
        Q_1&\!: z_1^3+z_2^3+z_3^3+z_4^3=0 \\
        Q_2&\!: w_1^3+w_2^3+w_3^3+w_4^3=0 \\
        Q_3&\!: \sum z_iw_i=0 \\
    \end{aligned}$};

    \draw[-] (z1) -- (w1);
    \draw[-] (z2) -- (w2);
    \draw[-] (z3) -- (w3);
    \draw[-] (z4) -- (w4);
    \draw[-] (z1) -- (z2);
    \draw[-] (z4) -- (z3);
    \draw[-] (w1) -- (w2);
    \draw[-] (w4) -- (w3);
    \draw[-] (z1) -- (z4);
    \draw[-] (z1) -- (z3);
    \draw[-] (z2) -- (z3);
    \draw[-] (z2) -- (z4);
    \draw[-] (w1) -- (w4);
    \draw[-] (w1) -- (w3);
    \draw[-] (w2) -- (w3);
    \draw[-] (w2) -- (w4);
    
    \draw[double,<->] ($(mid.east) + (-1,0)$) -- ($(mid.west) + (+1,0)$) node[midway,above] () {$S_4\times\mathbb{Z}_2$};

\end{tikzpicture}      
    \caption{The symmetry of a Tian-Yau manifold (right), and a matching graph (left). }
    \label{fig:tian_graph_sym}
\end{figure}
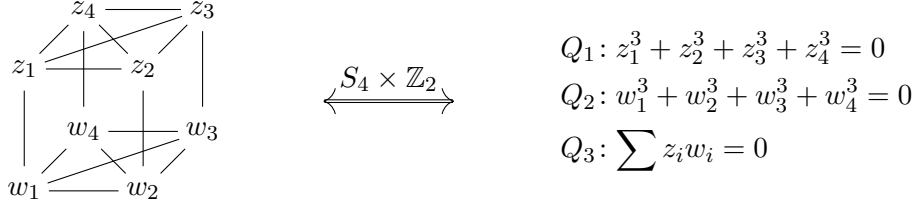

To encode permutation symmetries of Calabi--Yau coordinates we establish a connection between coordinates and graphs, which allows us to use GNNs instead of MLPs. For example, we can represent the permutation symmetries of K3 (\cref{fig:k3_graph_sym}) or Tian-Yau (\cref{fig:tian_graph_sym}) with appropriately chosen graphs. Better still, a suitable graph can be constructed for any symmetry group \cite{frucht_herstellung_1938}.

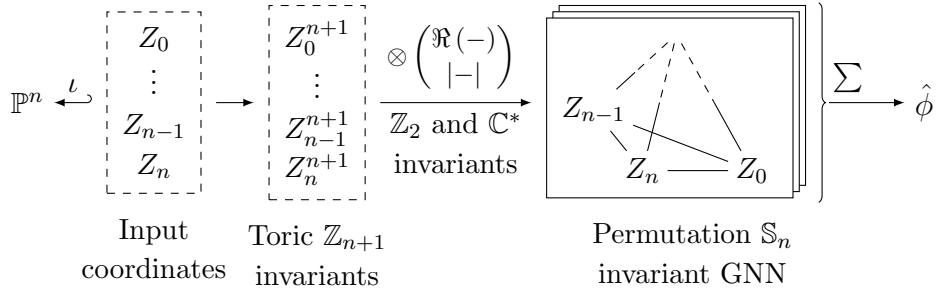
\begin{figure}[ht]
    \centering
    \begin{tikzpicture}[every text node part/.style={align=center}]
    \tikzstyle{box}=[rectangle, draw=black, inner sep = 2mm, outer sep = 2mm]
    \tikzstyle{dcs}=[double copy shadow, shadow xshift=2pt, shadow yshift=-2pt, fill=white, yshift=-2pt]
    
    \node[box, dashed] (coords) {$Z_0$ \\ $\vdots$ \\ $Z_{n-1}$ \\ $Z_n$};
    \node[box, dashed] (toric) [right = 0.5 of coords] {
        $Z_0^{n+1}$ \\ $\vdots$ \\ $Z_{n-1}^{n+1}$ \\ $Z_n^{n+1}$
    };
    \node[box, dcs] (gnn) [right = 2 of toric] { 
        \begin{tikzpicture}[every node/.style={inner sep = 0mm, outer sep = 1mm}]
            \node (z1) at (0.8, -0.8) {$Z_0$};
            \node (zd) [left = 0.8 of z1] {$Z_n$};
            \node (zn) at (-1.5, 0) {$Z_{n-1}$};
            \node (others) at (0, 1) {};
            
            \draw[-] (z1) -- (zd);
            \draw[-] (z1) -- (zn);
            \draw[-] (zd) -- (zn);
        
            \draw[-] (z1) -- ($(others)!0.5!(z1)$);
            \draw[dashed] (others) -- ($(others)!0.5!(z1)$);
        
            \draw[-] (zd) -- ($(others)!0.5!(zd)$);
            \draw[dashed] (others) -- ($(others)!0.5!(zd)$);
        
            \draw[-] (zn) -- ($(others)!0.5!(zn)$);
            \draw[dashed] (others) -- ($(others)!0.5!(zn)$);
            
        \end{tikzpicture}
    };
    \node (ambient) [left = 0.5 of coords] {$\PP^n$};
    
    \node [below = 0 of coords] {Input \\ coordinates};
    \node [below = 0 of toric] {Toric $\mathbb{Z}_{n+1}$ \\ invariants};
    \node [below = 0 of gnn, xshift = 3mm] {Permutation $\SS_n$ \\ invariant GNN};

    \draw[left hook-latex] (coords) -- (ambient)  node[midway,above] () {$\iota$};
    \draw[-latex] (coords) -- (toric);
    \draw[-latex] (toric) -- ($(gnn.west)+(0,2pt)$) node[midway, above] (conj_inv_transform) {\small
        $\otimes \begin{pmatrix} \Re\left(-\right) \\ \abs{-} \end{pmatrix}$
    };
    \node [below = 0 of conj_inv_transform] {$\ZZ_2$ and $\CC^*$\\ invariants};

    \draw[decoration={brace},decorate] ($(gnn.north east) + (4pt,0)$) -- ($(gnn.south east) + (4pt, 4pt)$) node[midway] (sum_mid) {};

    \node (phi) [right = of sum_mid] {{\large $\hat\phi$}};
    \draw[-latex] (sum_mid) -- (phi) node[near start, above] () {$\sum$};
\end{tikzpicture}
    \caption{A complete pipeline that incorporates $\CC^*$, $\ZZ_2$, $\SS_n$, and $\ZZ_{n+1}^n$ symmetries into a NN model. Note that spectral features $Z_i\overline{Z_j}$ must be encoded into edge features of GNN layers in order to align with the permutation symmetries. }
    \label{fig:gnn_pipeline}
\end{figure}

Thus, a complete pipeline that incorporates all of these symmetries can look like \cref{fig:gnn_pipeline}. We obtain good results when training neural networks this way, and report the $\sigma$-measures in \cref{tab:phi_by_norms}.

Despite these benefits, we do not pursue this approach further for several reasons, and we defer refinements and investigations using GNNs to the future. The first reason is that finding optimal invariants is very hard in general, although much progress has been made for free group actions \cite{braun_free_2011}.
The second reason is that we will conjecture that Fermat Calabi--Yaus have much larger symmetries inherited from the embedding in $\PP^n$. This stands to significantly speed up machine learning approaches in highly symmetric situations. In contrast, the neural network and GNN approaches are more useful in less symmetric contexts, such as the Dwork family, where we do not find additional symmetries.

\subsection{Salience of Symmetric Features}

To estimate the impact of different features on the loss, we train a model that uses many features with different symmetry groups. We use spectral features of \textcite{berglund_machine_2023}, as well as other more symmetric (0,0)-forms, and then compare their saliencies $a_i$ defined in \cref{eq:salience}. The most symmetric expressions we look at are the power sum \emph{symmetric polynomials}
\begin{equation}\label{eq:symm_forms}
    s_k = \frac{\sum_i\abs{Z_i}^{2k}}{\left(\sum_i\abs{Z_i}^2\right)^k}
\end{equation}
which we normalise to ensure they are degree $0$ homogeneous. These are real polynomials, so they do \emph{not} form a basis for symmetric functions of type $\PP^n\to\RR$. But, they \emph{do} form a basis on $\RR\PP^n$, where we only need to use $s_2, \dots, s_{n+1}$ because we always have $s_0=n$ and $s_1=1$.

\begin{figure}[ht]
    \centering
    \begin{tikzpicture} [node distance=7mm]
    
    \node (coords) {Coords $z$};
    \node (feats) [below= of coords] {Features $x$};
    \node (g) [below= of feats] {Metric $g$};
    \node (attrib) [below= of g] {Salience $a$};

    \draw[<-] (feats) -- node[midway,left] {$f(-)$} (coords);
    \draw[<-] (g) -- (feats) node[midway,left] {$\iota^*(g\FS+\dd \phi(-))$} (coords);
    \draw[<-] (attrib) -- (g) node[midway,left] {$\abs*{\partial\mathcal{L}(-)/\partial x}$};
    
    \draw[<-] (g.east) to [out=60,in=-60] node[midway,right] {$\dd$} (coords.east);
    \draw[red,<-] (attrib.east) to [out=60,in=-60] node[midway,right] {$\partial$} (feats.east);

    \node (coords2) [right= 4cm of coords] {Coords $z$};
    \node (feats2) [below= of coords2] {Features $x$};
    \node (g2) [below= of feats2] {Metric $g$};
    \node (attrib2) [below= of g2] {Salience $a$};

    \draw[<-] (feats2) -- node[midway,left] {$f(-)$} (coords2);
    \draw[<-] (g2) -- (feats2) node[midway,left] {$\iota^*(g\FS+ f^* \dd \phi(-))$};
    \draw[<-] (attrib2) -- (g2) node[midway,left] {$\abs*{\partial\mathcal{L}(-)/\partial x}$};
    
    \draw[<-, transform canvas={xshift = 3mm}] (g2.north) to [out=45,in=-45] node[midway,right] {$\dd$} (feats2.south);
    \draw[<-] (attrib2.east) to [out=60,in=-60] node[midway,right] {$\partial$} (feats2.east);
  
\end{tikzpicture}    
    \caption{Computation graph of attribution weights, with arrows denoting dependencies. Left: differentiation with respect to features is impossible as it overlaps with an earlier one. Right: correct implementation with both differentiations in feature space.}
    \label{fig:attrib_graph}
\end{figure}
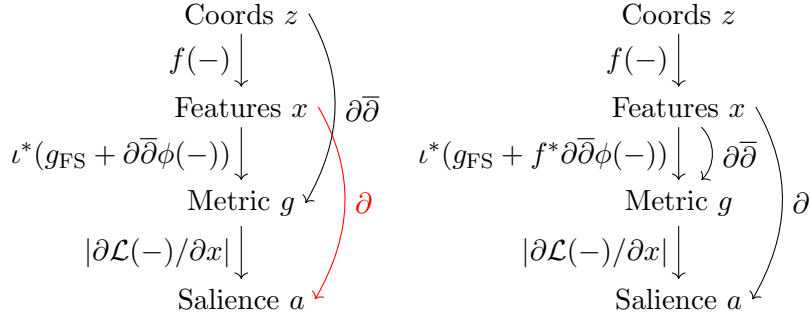

The saliencies $a_i$ cannot be computed in the PhiModel, as $g$ and $\mathcal{L}_\sigma$ are obtained by differentiating with respect to coordinates $Z_i$, while features $x=f(z,\overline{z})$ are in the middle of the computation graph and so are unavailable for differentiation. In order to compute them, we modify the PhiModel of \cref{eq:fs_corr} and replace differentiation $\partial\phi/\partial z$ in ambient space with differentiation $\partial\phi/\partial x$ in feature space via the chain rule in \cref{sec:chain_rule}. This approach is visualised in \cref{fig:attrib_graph}, and the resulting ansatz for the metric becomes
\begin{equation}
    g = \iota^*\paren*{
    g\FS + \frac{\partial \phi}{\partial x_k} \frac{\partial^2 x_k}{\partial z_i \partial \overline{z_j}} + \frac{\partial^2 \phi}{\partial x_k \partial x_l} \frac{\partial x_k}{\partial z_i} \frac{\partial x_l}{\partial \overline{z_j}}
    } 
\end{equation}
where $\phi:\RR^m\to\RR$ is now a function of real features $x(z,\overline{z})$, and $x: \PP^n\to\RR^m$. Unlike \cref{eq:fs_corr}, this expression \emph{can} be differentiated with respect to $x$, because we can treat $\partial x/\partial z$ terms as constants, and only differentiate the $\partial \phi/\partial x$ terms. This also means that we can theoretically improve backpropagation, by precomputing the pullback to features and differentiating only up to $x$. We do not implement this optimisation, as the training loop is already sufficiently fast for our purposes.

In our experiment we use the same architecture and parameters as described in \cref{sec:phi_conjecture}, with added weight decay of $\num{5e-4}$ to slightly optimise towards feature sparsity. We use random noise and 0 as baselines. Due to stochasticity, even the salience of a constant zero feature will never be zero --- we observe that it is just above $10^{-5}$. Notably, a constant non-zero feature would not be a valid baseline, as it would have much larger salience due to being equivalent to a bias term in the neural network.

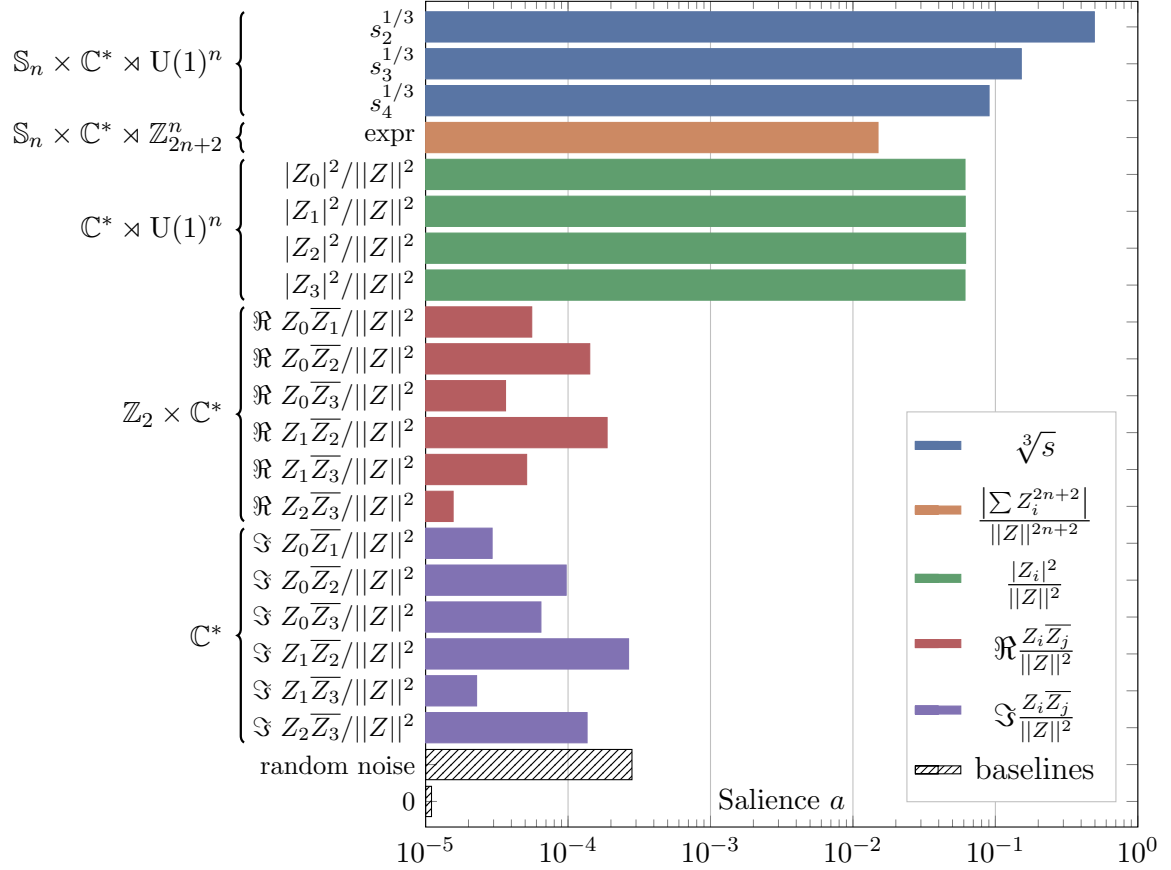
\begin{figure}[ht]
    \centering
    \pgfplotstableread{
value label
4.95e-01 1
1.52e-01 2
9.03e-02 3
}\attribblue

\pgfplotstableread{
value label
1.50e-02 4
}\attriborange

\pgfplotstableread{
value label
6.12e-02 5
6.14e-02 6
6.18e-02 7
6.12e-02 8
}\attribgreen

\pgfplotstableread{
value label
5.56e-05 9
1.42e-04 10
3.64e-05 11
1.88e-04 12
5.12e-05 13
1.56e-05 14
}\attribred

\pgfplotstableread{
value label
2.93e-05 15
9.70e-05 16
6.45e-05 17
2.66e-04 18
2.28e-05 19
1.36e-04 20
}\attribpurple

\pgfplotstableread{
value label
2.81e-04 21
1.10e-05 22
}\attribblack

\newcommand*{\xmin}{10^-5}

\begin{tikzpicture}
    \pgfplotsset{    
        width=11cm,
        height=12.5cm,
        every axis y label/.style={
            at={(current axis.above origin)},
            anchor=north east,
        },
        x label style={at={(axis description cs:0.5,0.01)},anchor=south},
        xlabel={Salience $a$},
        every y tick label/.append style={font=\small}
    }
    \begin{axis}[
        log origin=infty,
        xmode=log, 
        xmin=\xmin, xmax=1,
        xmajorgrids=true,
        clip=false,
        enlarge y limits=0.033,
        y dir=reverse,
        legend pos=south east, 
        legend style={draw=gray!50, 
                        font=\large, 
                        nodes={inner sep=5pt}},
        bar width=0.4cm,
        ytick={1,2,...,22},
        yticklabels={$s_2^{1/3}$,
                         $s_3^{1/3}$,
                         $s_4^{1/3}$,
                         expr,
                         $|Z_0|^2/||Z||^2$,
                         $|Z_1|^2/||Z||^2$,
                         $|Z_2|^2/||Z||^2$,
                         $|Z_3|^2/||Z||^2$,
                         $\Re\ Z_0\overline{Z_1}/||Z||^2$,
                         $\Re\ Z_0\overline{Z_2}/||Z||^2$,
                         $\Re\ Z_0\overline{Z_3}/||Z||^2$,
                         $\Re\ Z_1\overline{Z_2}/||Z||^2$,
                         $\Re\ Z_1\overline{Z_3}/||Z||^2$,
                         $\Re\ Z_2\overline{Z_3}/||Z||^2$,
                         $\Im\ Z_0\overline{Z_1}/||Z||^2$,
                         $\Im\ Z_0\overline{Z_2}/||Z||^2$,
                         $\Im\ Z_0\overline{Z_3}/||Z||^2$,
                         $\Im\ Z_1\overline{Z_2}/||Z||^2$,
                         $\Im\ Z_1\overline{Z_3}/||Z||^2$,
                         $\Im\ Z_2\overline{Z_3}/||Z||^2$,
                         \text{random noise},
                         $0$
                         },
    ]

    \addplot+[
        xbar,
        mark=none,
        color=deepBlue,
        fill=deepBlue,
    ] 
    table [y=label, x=value, col sep=comma] {\attribblue};
    
    \addplot+[
        xbar,
        mark=none,
        color=deepOrange,
        fill=deepOrange,
    ] 
    table [y=label, x=value, col sep=comma] {\attriborange};
    
    \addplot+[
        xbar,
        mark=none,
        color=deepGreen,
        fill=deepGreen,
    ] 
    table [y=label, x=value, col sep=comma] {\attribgreen};

    \addplot+[
        xbar,
        mark=none,
        color=deepRed,
        fill=deepRed,
    ] 
    table [y=label, x=value, col sep=comma] {\attribred};

    \addplot+[
        xbar,
        mark=none,
        color=deepPurple,
        fill=deepPurple,
    ] 
    table [y=label, x=value, col sep=comma] {\attribpurple};

    \addplot+[
        xbar, 
        mark=none,
        pattern=north east lines,
        solid,
        draw=black,
    ] 
    table [y=label, x=value, col sep=comma] {\attribblack};
    
    \legend{$\sqrt[3]{s}$, 
    $\frac{ \left|\sum Z_i^{2n+2}\right|}{||Z||^{2n+2}}$, 
    $\frac{|Z_i|^2}{||Z||^2}$,
    $\Re \frac{Z_i \overline{Z_j}}{||Z||^2}$,
    $\Im \frac{Z_i \overline{Z_j}}{||Z||^2}$,
    baselines
    }

    \newcommand*{\bracexshift}{-24mm}
    
    \draw[decoration={brace},decorate,line width=0.3mm, xshift=\bracexshift] 
    (\xmin, 3.4) -- (\xmin, 0.6) 
    node[midway, left] () {$\SS_n \times \CC^* \rtimes \operatorname{U}(1)^n\ $};
    
    \draw[decoration={brace},decorate,line width=0.3mm, xshift=\bracexshift] 
    (\xmin, 4.4) -- (\xmin, 3.6) 
    node[midway, left] () {$\SS_n \times \CC^* \rtimes \ZZ_{2n+2}^n\ $};
    
    \draw[decoration={brace},decorate,line width=0.3mm, xshift=\bracexshift] 
    (\xmin, 8.4) -- (\xmin, 4.6) 
    node[midway, left] () {$\CC^* \rtimes \operatorname{U}(1)^n\ $};
    
    \draw[decoration={brace},decorate,line width=0.3mm, xshift=\bracexshift] 
    (\xmin, 14.4) -- (\xmin, 8.6) 
    node[midway, left] () {$\ZZ_2 \times \CC^*\ $};

    \draw[decoration={brace},decorate,line width=0.3mm, xshift=\bracexshift] 
    (\xmin, 20.4) -- (\xmin, 14.6) 
    node[midway, left] () {$\CC^*\ $};
    
    \end{axis}
\end{tikzpicture}
    \caption{Relative salience $a$ on K3 (log scale) on the test set. Larger score means higher attribution relative to other features so that total salience sums to $1$. Here $\CC^*$ represents projective scaling symmetry, $\ZZ_2$ is conjugation invariance, $\SS_n$ is the permutation group, $\ZZ_{n+1}^n$ is toric symmetry $(Z_j \mapsto e^{\frac{2i\pi}{n+1} k_j} Z_j)$, and $\operatorname{U}(1)^n$ is phase invariance $(Z_j \mapsto e^{i\phi_j} Z_j)$. }
    \label{fig:attrib_distrib}
\end{figure}

We plot the distributions of $a$ in \cref{fig:attrib_distrib} for the K3 manifold, scale them so that total salience is $1$, and observe that:

\begin{itemize}
    \item $\operatorname{U}(1)$ symmetric features have higher salience, with {\deepBlue \textbf{symmetric polynomials} $s_k$} having by far the highest effect on the loss, 
    \item {\deepGreen \textbf{diagonal spectral features} $\abs{Z_i}^2/\norm{Z}^2$} have equal attribution, suggesting the network learnt permutation symmetry of the coordinates,
    \item salience of {\deepRed \textbf{off-diagonal spectral features}} is two orders of magnitude below the next lowest feature, and on the same level as random noise.
\end{itemize}

For a similar analysis of features on a Fermat $3$-fold, see \cref{sec:quintic_attrib}.

One may ask if this can be explained by larger symmetry groups having more influence on the metric. However, this will not do for several reasons. Firstly, size of the symmetry group is not directly correlated with salience. Secondly, off-diagonal elements are not just less salient than the more symmetric features. Rather, they are used as much as the uniform random noise, which is to say not at all. Thirdly, $\operatorname{U}(1)^n$ is not a symmetry of the Calabi--Yau and does not preserve the defining polynomial, unlike its discrete subgroups --- toric and conjugation symmetries $\ZZ_{n+1}^n,\ZZ_2<\operatorname{U}(1)^n$. However, features with $\operatorname{U}(1)^n$ symmetry achieve a higher score nonetheless.

This raises the question of why symmetric polynomials $s_k$ perform so well on Fermat Calabi--Yaus, despite having \textquote{incorrect} symmetries.

\subsection{Extrinsic Symmetries of Calabi--Yau Hypersurfaces}
\label{sec:hidden_sym}

To answer this, and to state the main theoretical results of our paper, we will need to consider the relationship between two important scalar forms. The first is the defining polynomial $Q$ of the Calabi--Yau, which is well defined on the entire $\PP^n$. The second is the correction term $\phi$, which is a globally defined $0$-form on the Calabi--Yau. We wish to make a statement about symmetry groups $\Sym$ under which $Q$ and $\phi$ are invariant, i.e.\ groups where $g\in\Sym(f)$ iff $f(x)=f(g.x)$. It is already well known that
\begin{equation*}
    \Sym(\phi) = \Sym(Q) = \Isom(\mathrm{CY}, g\fl)
\end{equation*}
where $\Isom(\mathrm{CY}, g\fl)$ is the group of isometries (distance-preserving maps) of the metric on the manifold. It seems there is nothing else left to say, but there is.
\begin{conj}[Extrinsic Symmetries] \label{thm:hidden_sym}
    There exists a function $\phi^\PP: \PP^n\to\RR$ that extends $\phi$ from the Calabi--Yau to the ambient space, meaning $\phi^\PP\big\lvert_{\mathrm{CY}} = \phi$, and that satisfies
    \begin{equation*}
        \Sym(\phi^\PP) = \Sym(\norm{\nabla Q}) \geq \Sym(Q) = \Isom(\mathrm{CY}, g\fl)
    \end{equation*}
\end{conj}
Observe that $\Sym(\norm{\nabla Q}) \geq \Sym(Q)$ easily holds since every isometry of the latter also acts on the former. This can lead to new simplifications because $\Sym(\norm{\nabla Q})$ does not need to be upper-bound either by $\Isom(CY)=\Sym(Q)$, or even by $\Isom(\PP^n)=\operatorname{PU}(n)$. In the next section we will focus on families of Calabi--Yaus where this symmetry group is strictly larger than $\Isom\paren*{\mathrm{CY}}$, which will induce novel constraints on $\phi$. In doing so we will mainly use the following corollary.
\begin{prop}[Existence of Individual $U(1)$ Symmetry]
\label{prop:sledgehammer}
    Assume \cref{thm:hidden_sym}. If a variable $Z_i$ only appears as a monomial $Z_i^{n+1}$ in $Q$, then $\phi$ is invariant to changes in $\arg Z_i$ and depends only on $\abs{Z_i}$.
    \begin{proof}
        We can compute that in this case $\norm{\nabla Q}$ depends only on $\abs{Z_i}$ and not on $\arg Z_i$. This is equivalent to being $U(1)$ invariant with respect to $Z_i$, so this $U(1)$ symmetry propagates to $\phi$ according to \cref{thm:hidden_sym}.
    \end{proof}
\end{prop}
We note that the isometry group of the Calabi--Yau, $\Isom\paren{CY}$, is closely related to the Killing vector field, in that $\Isom\paren{CY}$ is continuous iff the Killing vector field exists \cite[Proposition 18.9.]{gallier_isometries_2020}. Therefore, if the Killing vector field vanishes then the manifold has no continuous symmetries. This is the case for the $3$-dimensional Calabi--Yaus, for example. Significantly, this is not in conflict with \cref{prop:sledgehammer}, because we are characterizing not $\Isom\paren{CY}$, but a potentially larger group $\Sym\paren{\phi^\PP}$.

We wish to emphasise that these statements should not be taken as a \textquote{final} theoretical explanation of this phenomenon. Rather, they are observations with great predictive power (\cref{sec:phi_fermat_nn,sec:01mixed}) that sometimes still fall short of capturing the full picture (\cref{sec:phi_dwork}).

\subsubsection{Relation to Integration Weights \texorpdfstring{$w$}{}}

We end this section by bringing out a curious property of the integration weights
\begin{equation}
\label{eq:integration_weights}
    w=\frac{\mathrm{dVol}_\Omega}{\mathrm{dVol}\FS}=\frac{\det g\fl}{\det \iota^* g\FS}
\end{equation}
The weights $w$ are independent of the coordinate system because the transformations of the two determinants cancel out under change of coordinates. Therefore, $w$ is a global $0$-form on the Calabi--Yau, although this is not immediately apparent from the definition since \cref{eq:integration_weights} can only ever be computed in local coordinates.
\begin{theorem}[Integration Weights Identity]
\label{thm:closed_form_w}
    Consider a Calabi--Yau hypersurface embedded into $\PP^n$ with the defining polynomial $Q$. Then, the integration weights $w$ satisfy
    \begin{equation*}
        w = \frac{\norm{Z}^{2n}}{\norm{\nabla Q}^2}
    \end{equation*}
    where $\norm{\cdot}$ is the quadratic norm of vectors of coordinates $Z$ and gradients $\nabla Q$.
    \begin{proof}
        See \cref{sec:proofs}.
    \end{proof}
\end{theorem}
This identity yields another formula for $w$, highlights the fact that it is a global $0$-form, and makes it easier to compute. Furthermore, it reveals that $w$ also has a natural extension to the ambient space through $\norm{\nabla Q}$. We find it interesting that $\norm{\nabla Q}$ reappears here, especially since both $\phi$ and $w$ in some sense quantify the difference between flat and Fubini--Study metrics.

\cref{thm:closed_form_w} can be generalised to CICY manifolds embedded into $\PP^n$ with multiple defining polynomials $Q_j$. However, in this case weights $w$ will depend not just on $\norm{Z_i}$ and $\norm{\nabla Q_j}$, but on more complex symmetric invariants also. We do not explore this further as we do not focus on CICYs in this paper.

\section{Neural Network Approximations to \texorpdfstring{$\phi$}{Phi}}
\label{sec:phi_conjecture}

In this section we study performance of neural networks when learning $\phi$ from the perspective of our framework. We show that \cref{prop:sledgehammer} has substantial experimental support, and that it can significantly simplify calculations in cases such as the Fermat family. We use the \emph{cymyc} library by \textcite{berglund_cymyc_2024} as it implements the training pipeline in JAX \cite{bradbury_jax_2018} and leverages JIT compilation, which we observe makes training much faster compared to alternatives written in TensorFlow. We use neural networks and not KANs or symbolic regression, because taking multiple symbolic derivatives in the training pipeline makes the latter impractically slow.

\subsection{The Fermat Family}
\label{sec:phi_fermat_nn}

On Fermat Calabi--Yaus all coordinates appear as monomials in $Q$ (\cref{eq:fermat_general_homo}). Thus, using \cref{prop:sledgehammer} (and assuming \cref{thm:hidden_sym}) we immediately arrive at the following result:

\begin{prop}
\label{prop:phi_by_norms}
    Assume \cref{thm:hidden_sym}. Then, for Fermat Calabi--Yaus defined by \cref{eq:fermat_general}, the correction $\phi$ from \cref{eq:fs_corr} is a real function of absolute values of coordinates:
    \begin{equation}
    \label{eq:phi_by_norms}
        \phi\left(z,\overline{z}\right)=f\left(\abs{z_1},\abs{z_2},\dots,\abs{z_n}\right)
    \end{equation}
\end{prop}

Therefore, with extrinsic symmetries from \cref{thm:hidden_sym} we are free to simplify our models of $\phi$ on Fermat manifolds by only using absolute values. One might ask if a similar result can be derived directly from $\Isom(\mathrm{CY})$ instead, but we show that this is not the case.

\begin{prop}
\label{prop:not_norms_in_general}
    Consider a $d$-dimensional Fermat Calabi--Yau and an arbitrary $0$-form $f:\mathrm{CY}\to\RR$ with the symmetry group $\Sym(f)=\Isom(\mathrm{CY})$. Then, every such $f$ can be written as $f(Z,\overline{Z})=g(\abs{Z_1},\dots,\abs{Z_n})$ for some $g:\RR^{d+2}\to\RR$ if and only if $d=1$.
    \begin{proof}
        See \cref{sec:proofs}.
    \end{proof}
\end{prop}

In conclusion, we postulate that $\phi$ does not behave as an arbitrary $0$-form. Instead, it acquires additional symmetries from its relationship to the flat metric, and so \cref{thm:hidden_sym} allows us to make non-trivial statements about it.

\subsubsection{Machine Learning \texorpdfstring{$\phi$}{phi} with ModNet}
\label{sec:phi_nn}

To test \cref{prop:phi_by_norms} computationally, we train a neural network that learns $\phi$ only from coordinates' absolute values. However, we obtain poor results with architectures like $\mathrm{MLP}\left(\abs{z_1}, \dots\right)$. This is because the flat metric depends on the \emph{derivatives} of $\phi$. Here the only non-trivial derivatives arise from the activation function, which limits model expressivity. Instead, we will use the symmetric polynomials $s_k$ defined in~\ref{eq:symm_forms} reproduced below 
\begin{equation*}
    s_k = \frac{\sum_i\abs{Z_i}^{2k}}{\left(\sum_i\abs{Z_i}^2\right)^k}
\end{equation*}
Our model, ModNet, becomes
\begin{equation}
\label{eq:mlp_by_norms}
    \hat\phi=\mathrm{MLP}\left(\sqrt[n+1]s_k^j\,, \text { for } j,k\in \left\{1,\dots,n+1\right\}
    \right)
\end{equation}
We use 2 layers of dimension 64 and GELU activation and skip connections. Because inputs depend only on absolute values $\abs{Z_i}$, so $\hat\phi$ too depends only on the absolute values. As the inputs are scale-invariant, $\hat\phi$ is a valid function on $\PP^n$ by construction. 

With our ML pipeline that uses $s_k$ we can improve upon previous approaches while being highly memory-efficient and training only for minutes on commercial GPUs. We also perform much better than the GNN pipeline we proposed earlier. We train for $5$ epochs with batch size of $4096$ (largest possible on our hardware) on a dataset of \num{e6} points. We use Adam with learning rate of \num{e-4} and initialise weights with variance of \num{e-4}.

\begin{table}[ht]
    \centering
    \caption{$\sigma$-measure for Fermat Calabi--Yaus. FS metric baseline is equivalent to setting $\phi=0$. We omit \cite{douglas_numerical_2021, douglas_numerical_2007, headrick_numerical_2005} as they do not evaluate the $\sigma$-measure.}
    \begin{tabular}{ccccc}
    \toprule
    \multicolumn{1}{c}{} & \multicolumn{4}{c}{Fermat CY} \\
    \cmidrule(rl){2-5}
         { } & {Torus} & {K3 } & {Quintic} & {Sextic} \\
         { Model } & {\cref{eq:torus}} & {\cref{eq:k3}} & {\cref{eq:quintic}} & {} \\
         
    \cmidrule(r){1-1} \cmidrule(rl){2-5}
        Fubini-Study metric ($\iota^* g\FS$) & \num{0.1015} & \num{0.2617} & \num{0.4985}  & \num{0.5217} \\
        
        \textcite{jejjala_neural_2021}\tablefootnote{Resulting metrics are not K\"ahler} & --- & \num{0.068} & \num{0.18} & --- \\
        \textcite{gerdes_cyjax_2022}\tablefootnote{Result inferred from Fig.\ 3} & --- & --- & \num{0.05} & --- \\
        \textcite{anderson_moduli-dependent_2021}\tablefootnote{Result inferred from Fig.\ 1} & --- & --- & \num{0.05} & --- \\
        \textcite{ashmore_machine_2020} & --- & --- & \num{0.021} & --- \\
        \textcite{douglas_numerical_2008} & --- & --- & \num{0.019} & --- \\
        \textcite{ek_calabi-yau_2024} & --- & --- & \num{0.01} & --- \\
        \textcite{larfors_learning_2021} & --- & --- & \num{0.0086} & --- \\
        \textcite{berglund_machine_2023}\tablefootnote{Result inferred from Fig.\ 31, and our own re-evaluation.} & --- & \num{4.2e-4} & \num{0.0030} & --- \\
        \textcite{hendi_learning_2024} & --- & --- & \num{0.0027} & --- \\
        \textcite{headrick_energy_2010}\tablefootnote{Result inferred from Fig.\ 15} & --- & --- & $\sim$\num{e-4} & --- \\
        
    \cmidrule(r){1-1} \cmidrule(rl){2-5}
        GNN, \cref{fig:gnn_pipeline} & \num{1.9e-04} & \num{8.8e-4} & \num{0.0119} & --- \\
        ModNet, \cref{eq:mlp_by_norms}  & \num{2.6e-05} & \num{3.7e-04} & \num{0.0010} & \num{0.0014} \\
        Distilled formula, \cref{eq:symbolic_ansatz}  & \num{3.6e-05} & \num{3.9e-04} & \num{0.0011} & --- \\
        Compressed formula, \cref{eq:phi_compressed}  & \num{4.4e-05} & \num{6.5e-04} & \num{0.0011} & ---\\
    \bottomrule
    \end{tabular}
    \label{tab:phi_by_norms}
\end{table}

Our result for the Quintic corresponds to the Donaldson's algorithm using global sections of degree $k=52$, as per the scaling law by \textcite{douglas_numerical_2008}. The best performance is still being achieved by \textcite{headrick_energy_2010} with a method that does not use machine learning. We also perform tests on higher-dimensional CYs ($d>4$), where our model behaves similarly. However, testing on larger dimensions exacerbates numerical issues, as the normalisation constant $\kappa$ from \cref{eq:ma} decays exponentially with dimension of Calabi--Yau.

We also test ModNet in the small architecture regime (2 hidden layers of size 32), and observe similar performance. Furthermore, we observe that the model converges to a minimum after just one epoch of training, and that longer training does not further reduce the loss. This is contrasted to spectral networks which have a more typical behaviour and can take up to $100$ epochs to converge. We conclude that the expressive power of ModNet should be attributed to our features $s_k$ and not to the architecture. Furthermore, since we are reaching a threshold with our model, this suggests that in order to achieve lower loss we should either reduce learning rate or switch to more stable optimisation methods such as \textcite{headrick_energy_2010}. We leave these for future work. 

As discussed in \cref{sec:background}, one can have metrics with low $\sigma$-measure but high curvature, and for this reason it is important to check other properties of the trained model also, including the Ricci curvature $R$ itself.

\begin{table}[ht]
    \centering
    \caption{Test measures on the Fermat Quintic \cref{eq:quintic}, estimated with MC integration on the manifold. Our ModNet beats previous ML approaches with respect to every metric.
    }
    \begin{tabular}{cccccc}
    \toprule
    \multicolumn{1}{c}{} & \multicolumn{5}{c}{Measure (test set)} \\
    \cmidrule(rl){2-6}
         { Model } & {$\sigma$-measure} & {$\mathcal{R}$-measure} & {$\max_\text{CY} c_3$} & {$\norm{R_{ij}}$} & {$\norm{G_{ij}}$} \\
         
    \cmidrule(r){1-1} \cmidrule(rl){2-6}
        FS metric ($\iota^* g\FS$) & \num{0.4985} & \num{0.2278} & \num{561.63} & \num{0.3511} & \num{179.00} \\
        Flat metric ($g\fl$) & $\mathbf{0}$ & $\mathbf{0}$ & unknown & $\mathbf{0}$ & $\mathbf{0}$ \\
        
        \textcite{berglund_machine_2023} & \num{0.0030} & \num{0.0083} & --- & \num{0.0161} & \num{12.2} \\
        \textcite{hendi_learning_2024} & \num{0.0027} & \num{0.0956} & --- & --- & --- \\
        
    \cmidrule(r){1-1} \cmidrule(rl){2-6}
        ModNet, \cref{eq:mlp_by_norms}  & \num{0.0010} & \num{0.0028} & \num{1.90} & \num{0.0039} & \num{2.25}\\
        Compressed \cref{eq:phi_compressed}  & \num{0.0011} & \num{0.0029} & \num{2.07} & \num{0.0045} & \num{0.93}\\
    \bottomrule
    \end{tabular}
    \label{tab:topo_inv_quintic}
\end{table}

\cref{tab:topo_inv_quintic} shows that our model achieves the most accurate predictions for a variety of different topological quantities, such as the Ricci tensor $R_{ij}$ and the Einstein tensor $G_{ij} = R_{ij} - \frac{1}{2} g_{ij} R$. In particular, while \textcite{hendi_learning_2024} observed an increase in $\mathcal{R}$-measure when encoding symmetries with fundamental domains, our ModNet does not exhibit this pathological behaviour.

All measures are averaged over the Calabi--Yau, except the middle column. It reports the largest value of the $3^\text{rd}$ Chern form $c_3$, which is defined using Chern--Weil theory as
\begin{equation}
    c_3(J) = i\frac{-2\tr(J^3) + 3 \tr(J^2)\tr(J) - \tr(J)^3}{48\pi^3}
\end{equation}
where $J$ denotes the curvature form associated with the metric $g$. The curvature form does not vanish even for the Ricci-flat metric, since the metric may not necessarily be Riemann-flat.

The top Chern form $c_3$ is also the Euler density of the manifold, so $\int c_3$ equals the Euler characteristic $\chi$, which is known to be $-200$ for the Quintic. Being a topological invariant, $\chi$ is always the same whichever metric one integrates. \textcite{berglund_machine_2023} observe $\chi\approx -196.43$ in their integrations, but this discrepancy is purely numerical. We integrate over $\num{5e6}$ points, which is $90$x more, and produce $\chi\approx -199.3\pm0.1$ with either flat, Fubini-Study, or even random metrics. Again, this is expected because \textquote{Chern classes are independent of the choice of connection, but\ [...]\ different connections will lead to different representatives of the cohomology classes} \cite{bouchard_lectures_2007}, so $c_3(J_1) = c_3(J_2)+\partial\omega$ for any $J_1$, $J_2$. Calabi--Yau is compact, so $\int \partial\omega=0$, and we conclude that $\int c_3(J)$ is independent of the choice of $J$, which is to say that it is independent of the metric. 

On the other hand, $\max c_3$ is not a topological invariant, and it yields strikingly different values for Fubini--Study and flat metrics. The precise value of $\max c_3$ for the flat metric is unknown, but based on \cref{tab:topo_inv_quintic} we can conjecture that $c_3(J\fl)\leq0$. 

\begin{figure}[ht]
\begin{tikzpicture}
     \pgfplotsset{    
            width=\textwidth/2,
            height=4.8cm,
            every axis y label/.style={at={(current axis.above origin)},
            anchor=north east,
            yshift = 1cm,},
            xlabel style={at={(rel axis cs:0.5,-0.1)}, anchor=north},
            ylabel style={at={(rel axis cs:-0.2,0.4)}, rotate=90, anchor=east},
            ymode=log,
            legend style={draw=gray!50, 
                fill=white, 
                draw opacity=1,    
                nodes={inner sep=4pt, opacity=1, fill opacity=1}
                }, 
            legend image post style={opacity=1},
            every axis plot/.append style={thick}
        }
    \begin{groupplot}[
        group style={
          group size=2 by 1, 
          horizontal sep=1.75cm, 
        },
    ]

    \nextgroupplot[
        legend pos=north west,
        scaled x ticks = false,
        xmin=-1.65e4,xmax=1e3,
        ymin=3e-8,ymax=5e-3,
        xlabel={Euler density $c_3$},
        ylabel={Volume},
        log origin=infty,
    ]
        \addlegendentry{$g\FS$}
        \addplot[
            const plot, 
            color=deepBlue,
            name path=fs
        ] table [x=x, y=count, col sep=comma]  {external/c3fs_new.csv};

        \addlegendentry{$g\fl$}
        \addplot[
            const plot, 
            color=deepOrange,
            name path=flat
        ] table [x=x, y=count, col sep=comma]  {external/c3flat_new.csv};

        \path[name path=axis] (axis cs:0,1e-8) -- (axis cs:1,1e-8);
        \addplot [
            thick,
            color=deepBlue,
            fill=deepBlue, 
            fill opacity=0.15
        ]
        fill between[
            of=fs and axis,
        ];
        \addplot [
            thick,
            color=deepOrange,
            fill=deepOrange, 
            fill opacity=0.15
        ]
        fill between[
            of=flat and axis,
        ];
        

    \nextgroupplot[
        legend pos=north east,
        scaled x ticks = false,
        xmin=0,xmax=38,
        ymin=3e-6,ymax=0.5,
        xlabel={Kretschmann scalar $K$},
        log origin=infty,
    ]
        \addlegendentry{$g\FS$}
        \addplot[
            const plot, 
            color=deepBlue,
            name path=fs
        ] table [x=x, y=count, col sep=comma]  {external/kretschmann_fs.csv};

        \addlegendentry{$g\fl$}
        \addplot[
            const plot, 
            color=deepOrange,
            name path=flat
        ] table [x=x, y=count, col sep=comma]  {external/kretschmann_flat.csv};

        \path[name path=axis] (1,1e-6) -- (34,1e-6);
        \addplot [
            thick,
            color=deepBlue,
            fill=deepBlue, 
            fill opacity=0.15
        ]
        fill between[
            of=fs and axis,
        ];
        \addplot [
            thick,
            color=deepOrange,
            fill=deepOrange, 
            fill opacity=0.15
        ]
        fill between[
            of=flat and axis,
        ];
        






    \end{groupplot}
\end{tikzpicture}
\caption{Histograms of Euler density $c_3$ (left), and Kretschmann scalar $K$ (right), sampled over $\num{e6}$ points on the Fermat quintic. Contributions are scaled by integration weights $w$, and the $y$-axis is normalised so that the total volume is $1$.}
\label{fig:modnet_loss_curve}
\end{figure}
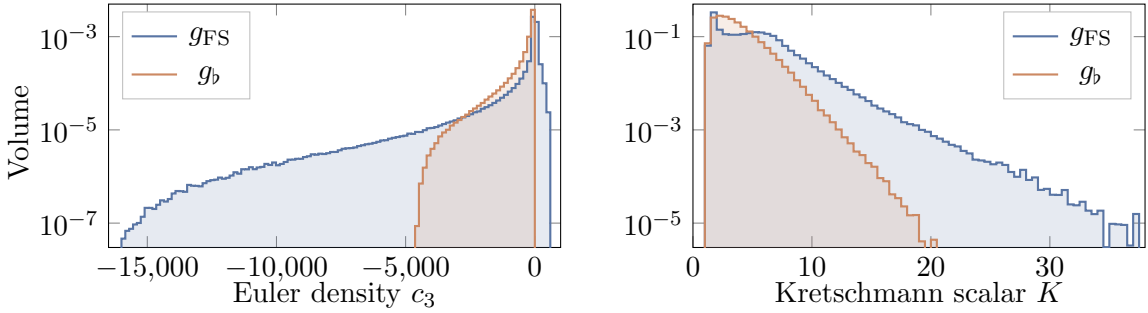

We also compute the Kretschmann scalar $K=\paren{R_{abcd}R^{abcd}}^{1/2}$ over the manifold. Since the Riemann tensor $R_{abcd}$ need not completely vanish, $K$ can take any positive real value. We plot the distributions of $c_3$ and $K$ in \cref{fig:modnet_loss_curve}. In both cases, these quantities have a smaller range with respect to the flat metric.
\subsection{The \texorpdfstring{$0$--$1$}{0--1} Mixed Family}
\label{sec:01mixed}

\begin{figure}[ht]
    \centering
    \pgfplotstableread{
value label
2.46e-01 0
1.36e-01 7
1.31e-01 12
1.25e-01 15
}\attribblue

\pgfplotstableread{
value label
1.84e-01 1
1.71e-01 2
}\attriborange

\pgfplotstableread{
value label
3.16e-04 3
1.99e-04 4
4.88e-04 5
2.17e-04 6
6.06e-04 8
8.52e-04 9
7.99e-04 10
6.20e-04 11
9.34e-04 13
1.38e-03 14
}\attribgreen

\newcommand*{\baseline}{5.85e-04}

\begin{tikzpicture}
    \pgfplotsset{    
        width=0.97\linewidth,
        height=6cm,
        every axis y label/.style={
            at={(current axis.above origin)},
            anchor=north east,
        },
        every y tick label/.append style={font=\tiny}
    }
    \begin{axis}[
        log origin=infty,
        xmode=log, 
        xmin=10^-4, xmax=1,
        xmajorgrids=true,
        enlarge y limits=0.04,
        y dir=reverse,
        legend style={draw=gray!50, 
                        font=\Large, 
                        row sep=4pt,       
                        fill=white, 
                        fill opacity=0.5, 
                        draw opacity=1,    
                        nodes={inner sep=4pt, opacity=1, fill opacity=1},      
                        at={(0.97,0.5)},
                        anchor=east
                        },
        legend image post style={opacity=1},
        bar width=0.2cm,
        ytick={0,1,...,15},
        yticklabels={$|Z_0|^2/||Z||^2$,
                         $\Re\ Z_0\overline{Z_1}/||Z||^2$,
                         $\Im\ Z_0\overline{Z_1}/||Z||^2$,
                         $\Re\ Z_0\overline{Z_2}/||Z||^2$,
                         $\Im\ Z_0\overline{Z_2}/||Z||^2$,
                         $\Re\ Z_0\overline{Z_3}/||Z||^2$,
                         $\Im\ Z_0\overline{Z_3}/||Z||^2$,
                         $|Z_1|^2/||Z||^2$,
                         $\Re\ Z_1\overline{Z_2}/||Z||^2$,
                         $\Im\ Z_1\overline{Z_2}/||Z||^2$,
                         $\Re\ Z_1\overline{Z_3}/||Z||^2$,
                         $\Im\ Z_1\overline{Z_3}/||Z||^2$,
                         $|Z_2|^2/||Z||^2$,
                         $\Re\ Z_2\overline{Z_3}/||Z||^2$,
                         $\Im\ Z_2\overline{Z_3}/||Z||^2$,
                         $|Z_3|^2/||Z||^2$
                         },
    ]

    \addplot+[
        xbar,
        mark=none,
        color=deepBlue,
        fill=deepBlue,
    ] 
    table [y=label, x=value, col sep=comma] {\attribblue};
    
    \addplot+[
        xbar,
        mark=none,
        color=deepOrange,
        fill=deepOrange,
    ] 
    table [y=label, x=value, col sep=comma] {\attriborange};
    
    \addplot+[
        xbar,
        mark=none,
        color=deepGreen,
        fill=deepGreen,
    ] 
    table [y=label, x=value, col sep=comma] {\attribgreen};

    \draw[dashed, ultra thick] (\baseline, -0.5) -- (\baseline, 15.5);
    
    \legend{$\frac{|Z_i|^2}{||Z||^2}$,
    $\frac{Z_0 \overline{Z_1}}{||Z||^2}$,
    $\frac{Z_i \overline{Z_j}}{||Z||^2}$} 

    \end{axis}

\end{tikzpicture}
    \caption{Salience of spectral features on a 2-dimensional 0--1 mixed CY ($n=3,\psi=2+i$). Black dashed line is the baseline salience of uniform random noise.}
    \label{fig:salience_12mix}
\end{figure}
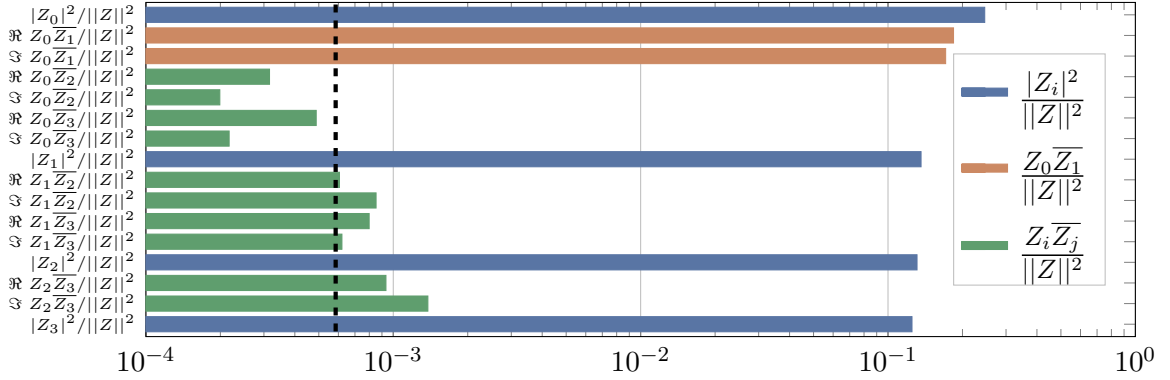

Now that we looked at the Fermat family in great detail, we will quickly go over some other symmetric Calabi--Yaus.  First, consider a family embedded into $\PP^n$ via
\begin{equation}
    Q_n\!:\ Z_0^{n+1}+Z_1^{n+1}+\dots+Z_n^{n+1} -\paren{n+1}\psi Z_0^n Z_1=0 
\end{equation}
which we name the 0--1 mixed family. Observe that this defining polynomial depends only on absolute values of each coordinate except $Z_0$ and $Z_1$. Therefore, \cref{prop:sledgehammer} predicts that $\phi$ can be parameterised using these $n+3$ terms only:
\begin{equation}
    \phi = f\paren*{\frac{\abs{Z_0}^2}{\norm{Z}^2}, \frac{\abs{Z_1}^2}{\norm{Z}^2},\cdots,\frac{\abs{Z_n}^2}{\norm{Z}^2}, \Re \frac{Z_0\overline{Z_1}}{\norm{Z}^2},\Im \frac{Z_0\overline{Z_1}}{\norm{Z}^2} }
\end{equation}
This is a theoretical prediction, and it is tentative as it requires assuming \cref{thm:hidden_sym}. However, it is in excellent agreement with experimental results, just like the Fermat family. In \cref{fig:salience_12mix} we indeed see that, just as we had predicted, only {\deepBlue \textbf{diagonal elements}} and {\deepOrange \textbf{off-diagonal elements that mix $Z_0$ and $Z_1$}} are salient. Meanwhile, salience of {\deepGreen \textbf{other diagonal elements}} is orders of magnitude smaller. In fact, it is consistent with salience of random noise, which we plot using a dashed black line. Thus, \cref{fig:salience_12mix} supports our theoretical result that these features are likely not used in the flat metric.

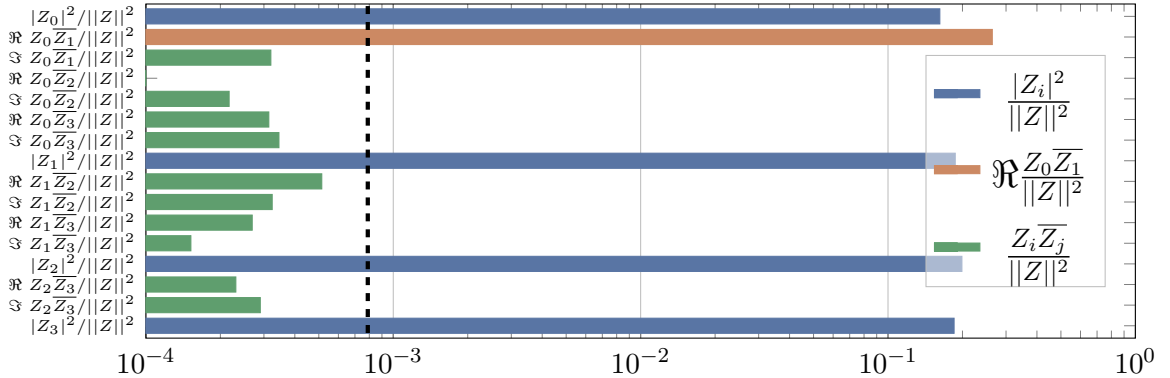
\begin{figure}[ht]
    \centering
    \pgfplotstableread{
value label
1.62e-01 0
1.87e-01 7
1.99e-01 12
1.85e-01 15
}\attribblue

\pgfplotstableread{
value label
2.64e-01 1
}\attriborange

\pgfplotstableread{
value label
3.20e-04 2
9.63e-05 3
2.17e-04 4
3.14e-04 5
3.45e-04 6
5.14e-04 8
3.24e-04 9
2.69e-04 10
1.52e-04 11
2.31e-04 13
2.90e-04 14
}\attribgreen

\newcommand*{\baseline}{7.89e-04}

\begin{tikzpicture}
    \pgfplotsset{    
        width=0.97\linewidth,
        height=6cm,
        every axis y label/.style={
            at={(current axis.above origin)},
            anchor=north east,
        },
        every y tick label/.append style={font=\tiny}
    }
    \begin{axis}[
        log origin=infty,
        xmode=log, 
        xmin=10^-4, xmax=1,
        xmajorgrids=true,
        enlarge y limits=0.04,
        y dir=reverse,
        legend style={draw=gray!50, 
                        font=\Large, 
                        row sep=4pt,       
                        fill=white, 
                        fill opacity=0.5, 
                        draw opacity=1,    
                        nodes={inner sep=4pt, opacity=1, fill opacity=1},      
                        at={(0.97,0.5)},
                        anchor=east
                        },
        legend image post style={opacity=1},
        bar width=0.2cm,
        ytick={0,1,...,15},
        yticklabels={$|Z_0|^2/||Z||^2$,
                         $\Re\ Z_0\overline{Z_1}/||Z||^2$,
                         $\Im\ Z_0\overline{Z_1}/||Z||^2$,
                         $\Re\ Z_0\overline{Z_2}/||Z||^2$,
                         $\Im\ Z_0\overline{Z_2}/||Z||^2$,
                         $\Re\ Z_0\overline{Z_3}/||Z||^2$,
                         $\Im\ Z_0\overline{Z_3}/||Z||^2$,
                         $|Z_1|^2/||Z||^2$,
                         $\Re\ Z_1\overline{Z_2}/||Z||^2$,
                         $\Im\ Z_1\overline{Z_2}/||Z||^2$,
                         $\Re\ Z_1\overline{Z_3}/||Z||^2$,
                         $\Im\ Z_1\overline{Z_3}/||Z||^2$,
                         $|Z_2|^2/||Z||^2$,
                         $\Re\ Z_2\overline{Z_3}/||Z||^2$,
                         $\Im\ Z_2\overline{Z_3}/||Z||^2$,
                         $|Z_3|^2/||Z||^2$
                         },
    ]

    \addplot+[
        xbar,
        mark=none,
        color=deepBlue,
        fill=deepBlue,
    ] 
    table [y=label, x=value, col sep=comma] {\attribblue};
    
    \addplot+[
        xbar,
        mark=none,
        color=deepOrange,
        fill=deepOrange,
    ] 
    table [y=label, x=value, col sep=comma] {\attriborange};
    
    \addplot+[
        xbar,
        mark=none,
        color=deepGreen,
        fill=deepGreen,
    ] 
    table [y=label, x=value, col sep=comma] {\attribgreen};

    \draw[dashed, ultra thick] (\baseline, -0.5) -- (\baseline, 15.5);
    
    \legend{$\frac{|Z_i|^2}{||Z||^2}$,
    $\Re \frac{Z_0 \overline{Z_1}}{||Z||^2}$,
    $\frac{Z_i \overline{Z_j}}{||Z||^2}$} 

    \end{axis}

\end{tikzpicture}
    \caption{Salience of spectral features on a 2-dimensional 0--1 mixed CY ($n=3,\psi=0.8$). Black dashed line is the baseline salience of uniform random noise.}
    \label{fig:salience_12mix_conj}
\end{figure}

Similarly, if we pick $\psi\in\RR$ then we obtain conjugation symmetry, which eliminates dependency on $\Im Z_0 \overline{Z_1}$ according to \cref{thm:hidden_sym}. Again, we observe this in practice in \cref{fig:salience_12mix_conj}, where most features are used as much as random noise. In particular, the $\Im Z_0 \overline{Z_1}$ feature that was highly salient in \cref{fig:salience_12mix} is almost completely unused here. 

We have to draw attention to a subtlety regarding the conjugation symmetry. Namely, if $\psi\in\RR$ then one can easily deduce that the defining polynomial $Q$ is conjugation invariant, and from this conclude that $\phi$ should be invariant under the map $Z\mapsto \overline{Z}$. Then the discussion in \cref{sec:encoding_symmetries} implies that one should use conjugation invariant features $\abs{\Im~Z_i\overline{Z_j}}$, and this indeed is the best possible feature to use assuming no other constraints are known. However, what we observe in \cref{fig:salience_12mix_conj} is not just that $\phi$ only depends on $\abs{\Im~Z_i\overline{Z_j}}$ instead of on $\Im~Z_i\overline{Z_j}$, but that $\phi$ \emph{does not depend on it at all}, a result that is only accessible through \cref{thm:hidden_sym}.

\subsection{The Dwork Family}
\label{sec:phi_dwork}

We also look at the Dwork family, defined via
\begin{equation}
    Q_n\!:\ Z_0^{n+1}+Z_1^{n+1}+\dots+Z_n^{n+1} -(n+1)\psi \prod Z_i=0 
\end{equation}
Note that we choose to scale the multiplicative term by $n+1$ for later convenience. This family is a non-example for our conjecture, because
\begin{equation*}
    \norm{\nabla Q_n}^2=(n+1)^2\sum_i \abs{Z_i^n-\psi \prod_{j\neq i}Z_j^2}
\end{equation*}
and we fail to find symmetries that were not already present in $Q_n$. Thus we cannot immediately deduce which reduced set of features to use --- \cref{thm:hidden_sym} is not applicable, so we would depend on $2n+2$ real and imaginary parts of coordinates. However, we empirically observe that we can train a neural network that matches spectral networks but uses only $n+1$ features: symmetric polynomials $s_2$ to $s_{n+1}$, and $w$. We take this to mean that deforming the Calabi--Yau with $\prod Z_i$ breaks symmetries in a way that requires including one more parameter. Furthermore, we note that this does not trivially generalise: features $s_2$ to $s_{n+1}$ and $w$ are not enough to learn the flat metric on other highly symmetric Calabi--Yaus, such as the Cefal\'u quartics \cite{catanese_kummer_2021, berglund_machine_2023} defined by 
\begin{equation*}
    Q\!:\ Z_1^4+Z_2^4+Z_3^4+Z_4^4 -\frac{\lambda}{3}\paren{Z_1^2+Z_2^2+Z_3^2+Z_4^2}^2 = 0
\end{equation*}
Thus this behaviour anticipates a finer handling of symmetries than what is done in this paper, and we hope that it will inspire future work. For now, we return to the Fermat Calabi--Yaus as we already understand them well enough, and there show that our symmetry arguments allow us to move from neural networks into the realm of closed form expressions.

\section{Symbolic Approximations to \texorpdfstring{$\phi$}{Phi}}
\label{sec:phi_symb}

In this section, we demonstrate how the black-box NN approximations can be distilled into interpretable formulae, \emph{with negligible impact to loss}. Specifically, we distill our approximations for $\phi$ on Fermat Calabi--Yaus into short expressions involving symmetric polynomials $s_k$. 

We will often focus on $3$-folds since they represent the extra dimensions in the 10-dimensional heterotic string theory. However, we are interested in the entire Fermat family, in order to show that it exhibits patterns in its flat metric which are generalisable across dimensions $d$.

\begin{figure}[ht]
    \centering
    \includegraphics[width=\textwidth]{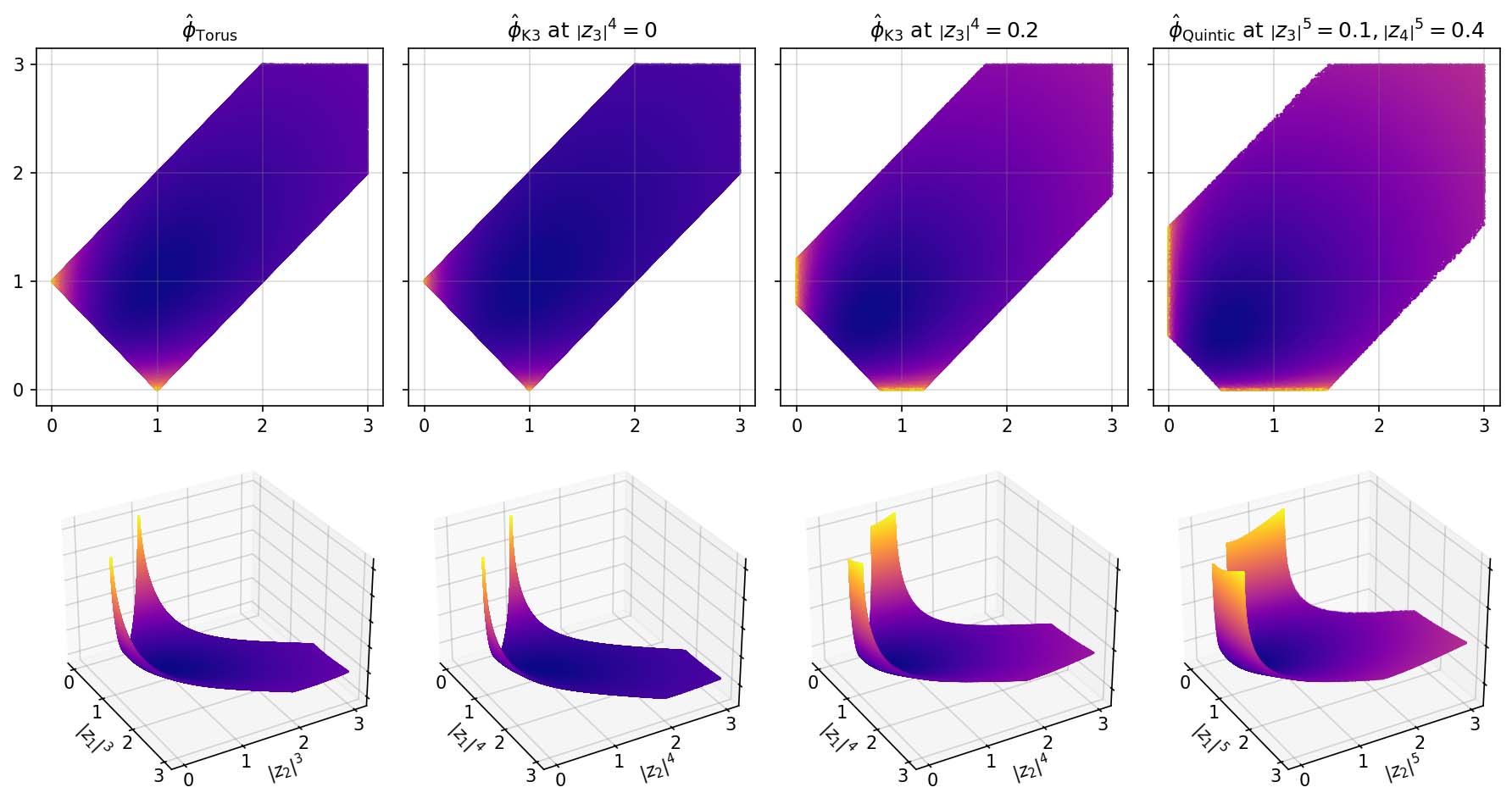}
    \caption{Leftmost: $\hat\phi_\mathrm{Torus}$ versus $\abs{z_1}^3$ and $\abs{z_2}^3$. Middle: $\hat\phi_\mathrm{K3}$ versus $\abs{z_1}^4$ and $\abs{z_2}^4$ for fixed $\abs{z_3}$. Rightmost: $\hat\phi_\mathrm{Quintic}$ versus $\abs{z_1}^5$ and $\abs{z_2}^5$ for fixed $\abs{z_3}$ and $\abs{z_4}$. Coordinates are plotted on $x$ and $y$ axes, while $\hat\phi$ is plotted on the $z$-axis and used to colour the figures. Note the similarity between all figures, especially the two left ones.
    }
    \label{fig:phi_by_norms}
\end{figure}

We do not focus on the K\"ahler potential $K$ in this section, because formulae for $\phi$ can always be converted into formulae for $K$ as $K\fl=K\FS+\phi$, but the other direction is nearly impossible due to an implicit change in coordinates. An example of this is \cref{prop:flat_kahler_torus}. It gives a local K\"ahler potential for the flat metric on the torus, but a global formula for $\phi$ is still unknown and thus worth studying.
\begin{prop}
\label{prop:flat_kahler_torus}
    Consider a complex torus hypersurface, and the canonical coordinate chart with a local coordinate $x=Z_2/Z_1$, and $Z_3$ eliminated via $Q\paren*{Z_1,Z_2,Z_3}=0$. Then, one possible K\"ahler potential of the flat metric can be obtained as
    \begin{equation*}
        K\fl(x,\overline{x}) = \abs*{\int \Omega(x) dx}^2
    \end{equation*}
    On the torus given by $Q(Z_1,Z_2,Z_3)=Z_1^3+Z_2^3+Z_3^3=0$, we have $\Omega=e^{-\frac{2i\pi}{3}}\paren{1+x^3}^{-\frac{2}{3}}$, so 
    \begin{equation*}
        K\fl(x,\overline{x}) = \abs*{\operatorname{Beta}\paren*{-x^3;\ \frac{1}{3},\frac{1}{3}}}^2
    \end{equation*}
    where $\operatorname{Beta}(x;\ a,b)=\int_0^x t^{a-1} (1-t)^{b-1} dt$ is the incomplete Beta function.
\end{prop}
We begin our symbolic analysis by plotting in \cref{fig:phi_by_norms} the correlation between $\hat{\phi}$ as approximated by ModNet and absolute values $z_i$. There, $\hat{\phi}$ appears as a well-behaved function of its inputs, a notion we will formalise later. 

\begin{figure}[ht]
    \centering
    \includegraphics[width=\textwidth]{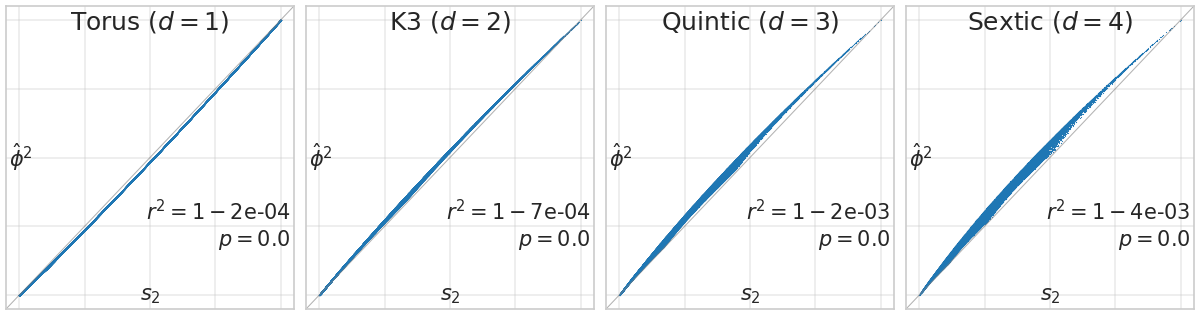}
    \caption{Correlation between $s_2$ ($x$-axis) and $\hat\phi^2$ ($y$-axis) for Fermat Calabi--Yaus. Axes have different scales. Important information is contained in the deviations from the straight line.}
    \label{fig:phi_vs_s2}
\end{figure}

Next, we plot $\hat{\phi}$ versus $s_2$ for multiple Calabi--Yaus in \cref{fig:phi_vs_s2}. We observe that this correlation is strong, but also non-linear. For the torus in particular, $\phi$ is very close but still not exactly equal to a scalar multiple of $s_2$. The nonlinearity for the torus is so small that it cannot be deduced from \cref{fig:phi_vs_s2}, nor with standard statistical tools. In fact, for all Calabi--Yaus we find that the $p$-value associated with a linear regression is so small that it physically cannot be stored as a floating point number, while the Pearson's correlation $r$ is consistently close to $1$. Thus, one needs to use different tools to explore these correlations. For example, on the torus one can find the expression of the metric obtained from $\phi=const \cdot s_2$ and show that it cannot be flat for any choice of $const$.

Our approach to studying $\hat\phi$ is to first distill the neural network into a symbolic form. We use the following ansatz, which we will motivate shortly:
\begin{equation}
\label{eq:symbolic_ansatz}
    \hat\phi_\mathrm{symb} = \sum_i\sum_j c_{ij}\sqrt[n+1]{s_i}^j
\end{equation}
We fit $c_{ij}$ via least squares optimisation from our ML approximated $\hat\phi$, and report the exact values in \cref{sec:coeffs}.

After distilling $c_{ij}$ from ML oracle we validate that the resulting formula achieves low $\sigma$-loss on test set, and report this in \cref{tab:phi_by_norms}. This validation step is important because a good approximation for $\phi$ does not have to approximate the metric $g$ at all. A model example is a piecewise linear function that can fit $\phi$ to arbitrary precision, but whose derivative vanishes everywhere. Thus, to achieve a good fit for $g$, we need an ansatz with \textquote{interesting} derivatives, and we discovered empirically that taking the roots in \cref{eq:symbolic_ansatz} maintains low loss after distillation. Specifically, this construction works markedly better than a power series in $s_k$. In a later section we will further explore the importance of taking the roots of these expressions.

To see how much information we lose when distilling and how much room for improvement there is, we compute $\abs{\hat\phi_\mathrm{ML}-\hat\phi_\mathrm{symb}}$ over the manifold and compare it to the machine epsilon for $32$-bit floating point numbers.

Machine epsilon $\epsilon=\num{1.192e-7}$ is not the smallest positive number that can be represented with $32$-bit floats under IEEE 754 \cite{noauthor_ieee_2019} --- that would be the subnormal $\num{1.4e-45}$. Rather, $\epsilon$ is the smallest positive number that numerically satisfies $1+\epsilon\neq 1$. The neural network is performing $32$-bit computation, and its inputs are $O(1)$, so machine epsilon is a good bound on the resolution of $\hat\phi$, and \cref{tab:distill_to_epsilon} shows that we are reaching the theoretical limit with our models.

\begin{table}[ht]
    \centering
    \caption{Difference between $\hat\phi_\mathrm{ML}$ and $\hat\phi_\mathrm{symb}$ in units of machine epsilon $\epsilon$.}
    \begin{tabular}{ccc}
    \toprule
    
    \multicolumn{1}{c}{} & \multicolumn{2}{c}{$\max_\text{CY}\abs{\hat\phi_\mathrm{ML} - \hat\phi_\mathrm{symb}}$} \\
    \cmidrule(rl){2-3} 
         { Calabi--Yau } & {Absolute} & {Relative to $\epsilon$} \\
         
    \cmidrule(r){1-1} \cmidrule(rl){2-2} \cmidrule(rl){3-3}
        K3 & \num{4.50e-07} & $\num{3.8}\epsilon$ \\
        Quintic & \num{1.93e-07} & $\num{1.6}\epsilon$ \\
    \bottomrule
    \end{tabular}
    \label{tab:distill_to_epsilon}
\end{table}

\subsection{Compressing Distilled Expressions}
\label{sec:interpretable}

We also generate interpretable formulae by interacting with PySR package for symbolic regression  \cite{cranmerDiscovering2020, cranmerInterpretableMachineLearning2023}. This way we discover a very compact approximation of the form 
\begin{equation}
\label{eq:phi_compressed}
    \hat\phi_\mathrm{symb} = \operatorname{poly}\paren{s_i}^\frac{1}{\pi}
\end{equation}
which we present in \cref{tab:phi_compressed}. Non-linear least-squares regression on polynomial coefficients converges in under a minute, and the end result has almost the same $\sigma$-loss as \cref{eq:symbolic_ansatz}, but $3$--$5$ times fewer parameters. For details behind this ansatz see \cref{sec:pysr}. In short, the exponent $\frac{1}{\pi}$ was determined empirically based on symbolic regression fits with PySR. This was performed on the torus, and then manually extended to higher dimensions. So, while it generalises well, it might not be locally optimal. This exponent could also be related to the same factor in the Fubini-Study potential, but due to lack of theoretical insight there is no obvious \emph{a priori} reason why one would choose one exponent over another, or one formula over another. 

\begin{table}[ht]
    \centering
    \caption{Compressed approximation for $\phi$ for different Fermat $d$-folds.}
    \begin{tabular}{cccc}
    \toprule
         {$d$} & {Symbolic $\hat\phi$} & {$\sigma$-measure} & { $\norm{R_{ij}}$} \\
    \midrule
         1 & $\left(\num{0.00501} + \num{0.01102} s_2 \num{-0.00157} s_3\right)^{\frac{1}{\pi}}$ & \num{4.44e-05} & \num{0.00164} \\
         2 & $\left(\num{0.00022} + \num{0.0068} s_2 + \num{0.00312} s_3 \num{-0.0012} s_4\right)^{\frac{1}{\pi}}$  & \num{6.46e-04}  & \num{0.0053}\\
         3 & $\frac{1}{5}\paren{\num{0.19} s_2 + \num{1.83} s_3 + \num{1.12} s_2^2 - \num{1.5} s_4 - \num{1.59}s_2s_3 + \num{1.72} s_5}^{\frac{1}{\pi}}$ & \num{0.0011} & \num{0.0045} \\
    \bottomrule
    \end{tabular}
    \label{tab:phi_compressed}
\end{table}

We fit these coefficients with least squares against the neural network $\hat\phi$ prediction. But, this becomes harder because \cref{eq:phi_compressed} is non-linear, and we cannot linearise it because $\phi$ is free up to a constant. These equations are also very sensitive to small changes in constants, which makes searching for them even more difficult.

\cref{tab:phi_compressed} shows that a more informed choice of ansatz to distill into can yield formulae with even fewer parameters than \cref{eq:symbolic_ansatz}. For example, for the Quintic we discover a formula with $\sigma$-loss that almost exactly matches the best neural network, but which only has \emph{six} free parameters to fit versus ten thousand. This again points to existence of significant structure in $\phi$. Approaches that are blind to this structure may therefore incur inefficiencies.

\subsection{Towards the Root of the Problem}

\begin{figure}[ht]
    \centering
    \includegraphics[width=\linewidth]{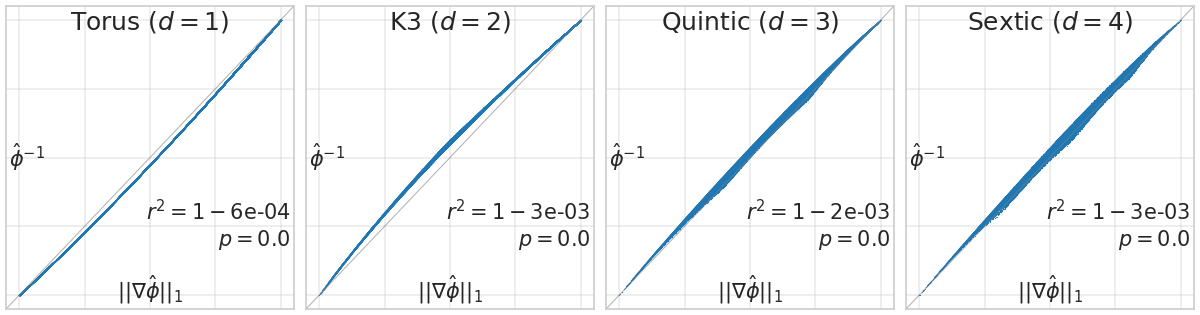}
    \caption{Correlation between $\hat\phi^{-1}$ and $\norm{\nabla \hat\phi}_1$ for Fermat Calabi--Yaus of dimensions $1$--$4$. Gradients are taken \emph{with respect to features} $x_i = \abs{Z_i}^2 / \norm{Z}^2$. Axes have different scales.}
    \label{fig:sum_of_grads_phi}
\end{figure}

Earlier we motivated the use of $n$-th roots in \cref{eq:phi_compressed} via their derivatives being \textquote{interesting}, but there is experimental motivation for this too. To see this, we first remark that $\phi$ (and our approximation $\hat\phi$) can be interpreted as a purely real function $\phi: \RR^n\to \RR$ with inputs $x_i=\abs{Z_i}^2/\norm{Z}^2$ according to \cref{prop:sledgehammer} and \cref{thm:hidden_sym}. Next, we study the correlation between $\hat\phi^{-1}$ and $\norm{\nabla \hat\phi}_1$ and visualise it in \cref{fig:sum_of_grads_phi}.

From this, we can empirically derive that on Fermat Calabi--Yaus $\phi^{-1} \sim \norm{\nabla \hat\phi}_1$.

\begin{prop}
    \label{prop:sqrt_ansatz}
    For $\alpha\in\RR^+$, $f:\RR^+\to \RR^+$ the functional equation
    \begin{equation*}
        \alpha f^{-n} = \norm*{\nabla f}_1
    \end{equation*}
    conditioned on $\sum x_i=1$, where $\norm{\cdot}_1$ is a L$^1$-norm, is solved by
    \begin{equation*}
        f=\paren*{\alpha \frac{n+1}{2}\sum x_i^2}^{\frac{1}{n+1}}
    \end{equation*}
    \begin{proof}
        One can compute $\frac{\partial}{\partial x_i}f = \alpha x_i f^{-n}$. Since the domain is $\RR^+$ we have $\norm{\nabla f}_1=\sum \frac{\partial}{\partial x_i}f$. Using $\sum x_i=1$ completes the proof.
    \end{proof}
\end{prop}

A relationship between $\hat\phi^{-1}$ and $\norm*{\nabla f}_1$ from \cref{fig:sum_of_grads_phi} thus produces an approximation $\phi\approx \sqrt{\alpha s_2}$, and is a possible underlying cause of the correlation between $\hat\phi^2$ and $s_2$ visible in \cref{fig:phi_vs_s2}. We say that the functional equation is \textquote{deeper} than this approximation, because the correlation between $\hat\phi$ and $s_2$ can be observed directly from \cref{fig:phi_vs_s2}, but in order to apply the functional equation from \cref{prop:sqrt_ansatz} and obtain \cref{fig:sum_of_grads_phi}, one first requires a specific parameterisation of $\hat\phi$ in terms of absolute values and this in turn requires our symmetry arguments and \cref{thm:hidden_sym}.

Consider now what this means for $\phi$. We found a PDE that it approximately satisfies, found a solution for it, and showed that it closely matches $\hat\phi$ itself back in \cref{fig:phi_vs_s2}. Now, $\phi$ is by definition a solution to a PDE, namely Monge--Amp\`ere \cref{eq:ma}, which begs the question whether there exists a \emph{different}, more solvable PDE that it also satisfies. Unfortunately, in the present case the formula $c \sqrt{s_2}$ is the worst performing distillate so far even though it approximates $\hat\phi$ well, but after slight modifications we arrive at \cref{eq:phi_compressed} which has the best ratio of parameters to accuracy.

\begin{figure}[ht]
    \centering
    \begin{tikzpicture}
    \pgfplotsset{    
        width=\linewidth,
        height=0.4\linewidth,
        every axis y label/.style={
            at={(current axis.above origin)},
            anchor=north east,
            yshift = 1cm,
        },
    }
    \begin{axis}[
        legend pos=north east,
        legend style={draw=gray!50},
        scaled y ticks = false,
        y tick label style={
            /pgf/number format/fixed,
            /pgf/number format/1000 sep = \thinspace
        },
        xmin=0.5,xmax=10^4,
        ymin=10^-4,ymax=1,
        xlabel={\# parameters},
        xlabel style={at={(rel axis cs:0.8,0)}, anchor=south west},
        ylabel={$\sigma$-measure},
        ylabel style={at={(rel axis cs:0,0.85)}, anchor=north west},
        ymode=log,
        xmode=log,
        ymajorgrids=true, 
        xmajorgrids=true,
        axis y line*=left, 
        axis x line*=none,
        nodes near coords,
        point meta=explicit symbolic,
        every node near coord/.style={
            font=\small,
            anchor=west,
            yshift=5pt,
            xshift=-2pt,
            color=black
        },
        minor xtick={1,2,...,10,20,30,...,100,200,300,...,1000,2000,3000,...,10000},
        extra x ticks={0.5},
        extra x tick labels={$0^{\vphantom{0}}$},
    ]
        \addlegendentry{Quintic}
        \addplot[deepBlue, mark=*] table [x=x, y=y, meta=label, col sep=comma] {pareto.csv};
        
        \addlegendentry{K3}
        \addplot[deepOrange, mark=*] table [x=x, y=y, col sep=comma] {pareto_k3.csv};

        \addplot[scatter, only marks, deepBlue, mark=o, thick] table [x=x, y=y, meta=label, col sep=comma] {extra.csv};
        \addplot[scatter, only marks, deepOrange, mark=o, thick] table [x=x, y=y, meta=label, col sep=comma] {extra_k3.csv};
    \end{axis}

    \draw (0.45,0.25) -- (0.45,-0.25); 
    \draw (0.25,0.25) -- (0.25,-0.25); 

\end{tikzpicture}
    \caption{Pareto frontier of multiple symbolic approximations for $\phi$, trading off number of parameters ($x$-axis) and $\sigma$-measure ($y$-axis). Filled points mark the frontier while empty points represent suboptimal results. Both axes are in $\log$ scale. Mind the discontinuity on the $x$-axis between $0$ and $1$. Near the top left \textquote{$\paren{\dots}$} represents an expression with no parameters that is defined in \cref{prop:zero_param_fit}, while constants in other formulae were fit numerically.}
    \label{fig:pareto_ansatz}
\end{figure}
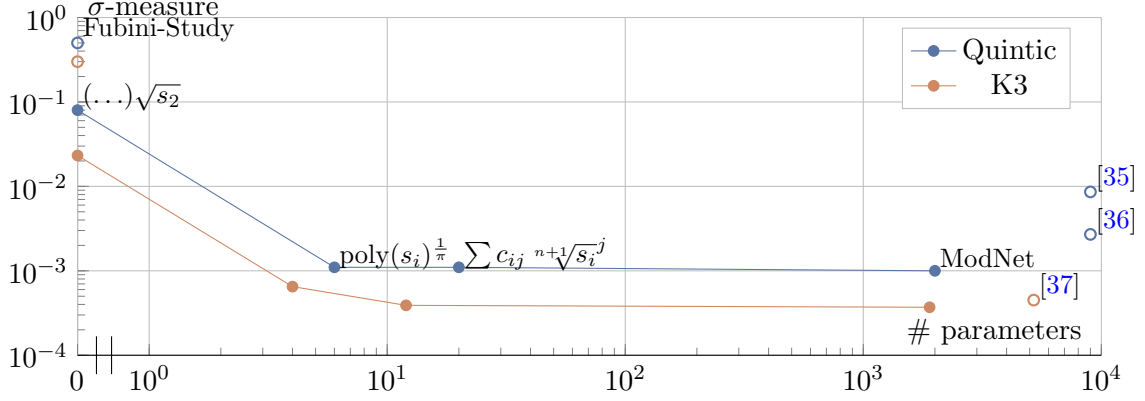

\cref{fig:pareto_ansatz} shows that the over-parameterised regime of neural networks achieves the lowest loss, which is inversely proportional to the number of parameters. However, it also shows that we can have both low loss and few parameters if we design custom ans\"atze that explicitly perform more complex calculations.

Note that a hypothetical closed-form expression for the flat metric would lie infinitely far on the bottom left of \cref{fig:pareto_ansatz}, because it would have no free parameters (for a fixed choice of complex structure moduli) and zero $\sigma$-measure because it is flat.

\section{Analytical Properties of \texorpdfstring{$\phi$}{Phi}}

So far we have shown how to use various symmetries of our learnt ML approximations to construct simple ans\"atze for nearly-flat metrics. This still does not explain how these relations arise in the first place, and whether they have any special meaning in the context of flat metrics. We concluded the last section by showing that one can find PDEs that yield good approximations to $\phi$ which could explain some of its behaviour. We are therefore motivated to investigate whether $\phi$ has other interesting analytical properties as well. In this section we thus turn to treating our approximations $\hat\phi$ more theoretically, as opposed to numerical experiments in previous sections. Firstly, we informally remark that our approximations $\hat\phi$ exhibit a curious pattern that is only visible \emph{across multiple dimensions}:

\begin{conj}[Informal]
\label{conj:self_similarity}
    On a $d$-dimensional Fermat Calabi--Yau, the correction $\phi$ is \textquote{related} to the correction one dimension up, with one coordinate zeroed out:
    \begin{equation}
        \phi_d\left(Z_0,\dots\right) \sim \phi_{d+1}\left(0,W_1,\dots\right)
    \end{equation}
    where $Z$ is a point on the $d$-fold and $W$ is a related point on the $d+1$-fold with $W_0=0$.
\end{conj}

This \textquote{similarity across dimensions} can be seen in \cref{fig:phi_by_norms}, where the plots are qualitatively identical despite originating from different Calabi--Yaus, suggesting that $\phi$ has a significant amount of structure. 

We can partially explain this behaviour because $s_k\left(Z_0,\dots\right)=s_k\left(0,Z_0,\dots\right)$, but the exact relationship is unclear. We can also demonstrate this more directly.

Observe that we cannot directly compare the corrections $\phi$ of Calabi--Yaus with different dimensions, as they also have different defining polynomials and ambient spaces. However, if $\paren{Z_0,\dots, Z_{a+1}}$ is a point on Fermat Calabi--Yau of dimension $a$, then $\paren{Z_0^\frac{a}{b},\dots,Z_{a+1}^\frac{a}{b}, \underset{b-a}{\underbrace{0,\dots,0}}}$ is a point on Fermat Calabi--Yau of dimension $b>a$. Thus, this relation allows us to meaningfully link two different corrections on a subset of dimension $a$.

For example, take a Fermat $1$-fold defined with $Z_0^3+Z_1^3+Z_2^3=0$, and a Fermat $2$-fold defined with $W_0^4+W_1^4+W_2^4+W_3^4=0$. Then, one can associate to each point $\paren{Z_0,Z_1,Z_2}$ on the $1$-fold a point $\paren{0,Z_0^{3/4},Z_1^{3/4},Z_2^{3/4}}$, and it is easy to check that it indeed lies on the $2$-fold. We claim that values of $\phi$ at these points are highly correlated, even though they come from different Calabi--Yaus. 

As a side note, the association we present is not unique, because it involves taking a complex $n$-th root. Nevertheless, our claim is well-defined due to toric symmetries of $\phi$, which make it invariant to the choice of the root branch.

\begin{figure}[ht]
    \centering
    \includegraphics[width=\linewidth]{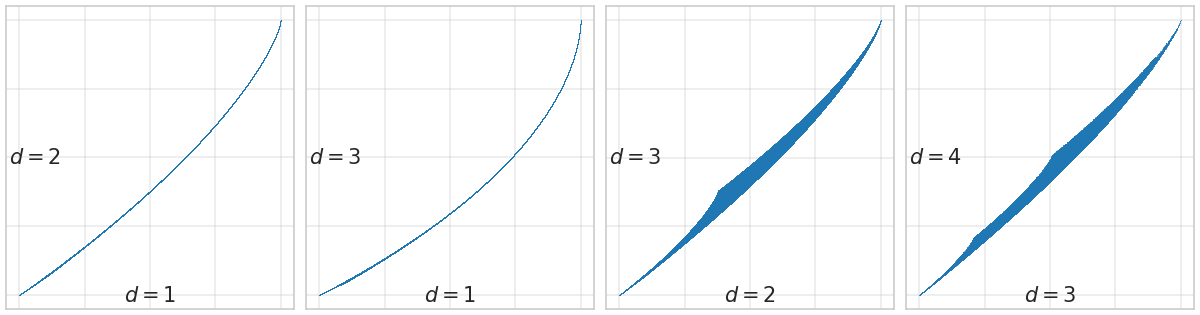}
    \caption{Correlation of $\hat\phi$ on different Fermat CYs. As explained above, smaller CYs' coordinates are power-scaled to satisfy the defining polynomial of larger CYs, and extra coordinates are set to $0$.}
    \label{fig:phi_substructure}
\end{figure}

In \cref{fig:phi_substructure} we plot such correlations for a range of different choices of dimensions. In each plot we again observe this \textquote{self-similarity} suggested by \cref{conj:self_similarity}. We also emphasise that both \cref{fig:phi_vs_s2,fig:sum_of_grads_phi}, and now also \cref{fig:phi_substructure}, share the same visual feature. Namely, scatter plots of $\hat\phi$ on CY of dimension $d$ seem to have $d-1$ sharp \textquote{corners} along it. It is surprising that this feature appears in all the plots, suggesting they may share a common explanation.

To make further statements about the analytic behaviour of $\phi$, we first define some terminology. Assume a coordinate chart on the CY under the standard pullback, where $Z_0=1$ and local coordinates are $Z_1\dots Z_n$. Then define the following loci on Fermat Calabi--Yaus.

Firstly, let the \emph{pseudo-origin} be the origin in this local coordinate chart. This point is not uniquely defined in the ambient space, but our observations hold for each of them due to symmetries. In ambient space, the pseudo-origin has coordinates $\paren{1,\sqrt[n+1]{-1},0,\dots,0}$ up to permutation.

Next, the \emph{equimodular locus} is the locus on Calabi--Yau defined by $\abs{Z_0}=\abs{Z_1}=\dots=\abs{Z_n}$. This submanifold has real dimension $d-1$, where $d=n-1$ is the complex dimension of the Calabi--Yau, because each of the $n$ constraints reduces dimension by one. Our investigation predicts that $\phi$ will be constant on this submanifold, so we can expect to see interesting simplifications on it.

These loci are well defined and non-empty for all Fermat hypersurfaces. We also propose that they are interesting to study because on them the behaviour of $\phi$ and $g\fl$ can be conjectured or outright proven.

\begin{conj}
    The correction $\phi$ attains a global maximum at any pseudo-origin, and a global minimum on the equimodular locus. More generally, if we fix absolute values for some $Z_i$ and vary the others, then the minimum is attained when the remaining coordinates have the same absolute value, and the maximum is attained when all but one of the free coordinates is zero.
\end{conj}

This conjecture says that $\phi$ is not just continuous and differentiable, but has sufficient analytic structure for stronger statements such as derivative bounds to be imposed on it.

\begin{conj}
    Consider a $d$-dimensional Fermat Calabi--Yau embedded into $\PP^n$ where $n=d+1$. Then there exist constants $a,b,c\in\RR$ with $0<a<b,~c$ such that the correction $\phi$ satisfies:
    \begin{align*}
        -\frac{c}{\abs{Z_i}^2} <& \frac{\partial_i \phi}{\overline{Z_i}} < \frac{c}{\abs{Z_i}^2} \\
        -\frac{a}{\abs{Z_i}^2} <& \partial_i\overline{\partial_i} \phi < \frac{b}{\abs{Z_i}^2} \\
        -\frac{b}{\abs{Z_i}^2\abs{Z_j}^2} <& \frac{\partial_i\overline{\partial_j} \phi}{\overline{Z_i}Z_j} < \frac{a}{\abs{Z_i}^2\abs{Z_j}^2} \\
    \end{align*}
    Note that $\partial_i \phi/\overline{Z_i} , \partial_i\overline{\partial_j} \phi/\overline{Z_i}Z_j \in \RR$ by \cref{thm:hidden_sym}, so the inequalities are well-posed. In particular, we conjecture that the exponent $2$ in the denominators is tight.
\end{conj}

It follows from this conjecture that $\dd \phi$ decays as $1/\abs{Z_i Z_j}$. Therefore, if we eliminate $Z_n$ via pullbacks, then the deviation from the Fubini-Study metric given by
\begin{equation*}
    \paren{\iota^* \dd\phi}_{ij}=\paren*{\partial_i - \frac{Z_i^n}{Z_n^n} \partial_n}\overline{\paren*{\partial_j - \frac{Z_j^n}{Z_n^n} \partial_n}}\phi
\end{equation*}
must be dominated by behaviour of $Z_i^n/Z_n^n$ terms in the limit of large $Z$, and not by the derivatives of $\phi$. One practical consequence would be that for points $Z$ with large $\norm{Z}$ but small $Z_i$ for some $i$, the flat metric along the direction $i$ looks like the Fubini--Study metric. This is consistent with our other observations, and also appears linked to \cref{conj:self_similarity}.

These statements and conjectures point to a deeper pattern in the structure of $\phi$ that we do not yet capture. We plan to revisit this curious behaviour in future work, and hope that by studying it we will learn more about properties of the Calabi--Yau flat metric.

\subsection{Exact Formulae for the Flat Metric}
\label{sec:exact_formulae}

In \cref{sec:phi_symb} we described how to find approximate expressions for $\phi$. Now, we ask if it is also possible to produce \emph{exact} formulae for the flat metric. To begin our search, we make the following assumption:

\begin{assumption}
\label{ass:closed_form}
    The flat metric may either be a closed form expression or an infinite series, but all numbers appearing in it must be computable up to a shared constant multiplicative factor.
\end{assumption}

By a \emph{computable number}, we mean that it can be independently computed to arbitrary precision and is not a result of, e.g., least squares regression on the formula. The flat metric is only unique up to scaling and we have to provision for this free factor in our assumption.

\cref{ass:closed_form} is strong and rules out all approximations from the previous section, meaning that a different approach is needed here. Due to the complexity of this problem, we do not hope to find a formula for the metric that works globally. Instead, we focus on specific loci where such a formula may be easier to find, such as the pseudo-origin $\paren{1,\sqrt[n+1]{-1},0,\dots,0}$ and the equimodular locus defined by $\abs{Z_1}=\abs{Z_2}=\dots=\abs{Z_n}$.
\begin{prop}
\label{prop:pseudo_origin}
    On a pseudo-origin of the Fermat Calabi--Yau, the flat metric $g\fl$ with the same volume as $g\FS$ satisfies
    \begin{equation*}
        g\fl\big|_{\paren{1,\sqrt[n+1]{-1},0,\dots,0}} = \lambda I
    \end{equation*}
    where $\lambda=\paren{\kappa/(n+1)^2}^{1/\paren{n-1}}$, $\kappa$ is the normalisation constant defined in \cref{eq:ma}, and $I$ is the identity matrix.
    \begin{proof}
        Consider the isometry generated by toric symmetries, and its associated Jacobian $J=\diag\paren{e^{\frac{2 a_k i \pi}{n+1}}}$ for arbitrary $a_k\in \ZZ$. Since the pseudo-origin is a fixed point of this isometry, we have $g\fl = J g\fl J^\dag$. This must hold for any choice of $a_k$, but one can show that this is possible only if $g\fl$ is a purely diagonal matrix here. Then, due to permutation symmetry, we deduce that it must be a scalar multiple of the identity matrix. The precise coefficient $\lambda$ is then derived from $\det g\fl=\kappa \abs{\Omega}^2$, because right-hand side simplifies to $\kappa/(n+1)^2$ on the pseudo-origin.
    \end{proof}
\end{prop}
\begin{prop}
\label{prop:equimodular}
    Assume \cref{thm:hidden_sym}. Then, the flat metric $g\fl$ of a Fermat Calabi--Yau with the same volume as $g\FS$, when restricted to the equimodular locus $X$, satisfies
    \begin{equation*}
        g\fl\big|_X = (n+1)\pi\lambda \iota^* g\FS
    \end{equation*}
    \begin{proof}
        Assuming that $\phi$ functionally depends only on absolute values, and applying the chain rule and permutation symmetry, we obtain $\dd\phi\vert_X= a I + b Z^\dag \otimes Z$. We obtain the desired expression by substituting this into $\det g\fl=\kappa\abs{\Omega}^2$ and following a derivation analogous to \cref{lemm:pullbacked_fs_det}.
    \end{proof}
\end{prop}
These propositions are significant because they give direct formulae for the elusive flat metric. The first is proven directly for a pseudo-origin of the Fermat Calabi--Yau, while the second requires our symmetry assumptions. Still, the second proposition is much stronger, as it says that we can fully describe the flat metric on a subspace of dimension approximately half that of the Calabi--Yau. 

The coefficient $\lambda$ plays the role of the free parameter in these equations, since $\lambda$ is precisely derived from fixing the volume of the metric. It is also noteworthy how cleanly this parameter is shared between the formulae, and that fixing the volume affects both formulae in \emph{exactly the same way}.

In addition to these two statements, we also present these conjectures about $g\fl$ and $\phi$:
\begin{conj}
    On a Fermat Calabi--Yau of dimension $d>1$, only the pseudo-origins and the equimodular locus satisfy the equation
    \begin{equation*}
        g\fl = \paren*{\frac{\det g\fl}{\det \iota^* g\FS}}^{1/\paren{n-1}} \iota^* g\FS
    \end{equation*}
\end{conj}
One can check that this equation is consistent with both \cref{prop:pseudo_origin,prop:equimodular}. 

Furthermore, it can be used to approximate the flat metric of nearby points due to continuity, but we cannot stretch this too far. Firstly, the formula is not K\"ahler in higher dimensions, so it cannot work globally except in dimension $1$ where it is trivially true for any Calabi--Yau. Secondly, the formula only provides solid estimate close to the equimodular locus and pseudo-origins. For example, on the Quintic, on a dataset of $15000$ points where all absolute values are within $10\%$ of each other ($\min \abs{Z_i} / \max \abs{Z_i} > 0.9$), we compute that the relative error $\norm{g\fl - g_\mathrm{approx}}_F / \norm{g\fl}_F$ never exceeds $6\%$, with a mean of just $2\%$, where $\norm{\cdot}_F$ is the Frobenius norm. 

On the other hand, far away from the equimodular locus the error increases to $42\%$. It is interesting that this error does not diverge globally even though it does not have a theoretically guaranteed upper bound, suggesting that this approximation performs better than a random expression would. Specifically, it performs better than a simpler estimate $g\fl\approx \iota^* g\FS$, for which the relative error reaches $79\%$ globally.

Still, there are better ways to generalise \cref{prop:pseudo_origin,prop:equimodular}, and in a way that at least produces K\"ahler metrics. We can take the previously established relation $\phi\approx \sqrt{\alpha s_2}$ and solve for $\alpha$ on the pseudo-origin.

\begin{prop}
\label{prop:zero_param_fit}
    A potential $\phi=\sqrt{\alpha s_2}$ yields the metric $\paren{\frac{1}{2\pi}-\frac{\sqrt{2}}{4}\alpha} I$ on the pseudo-origin. Thus, the choice of $\alpha$ that yields the flat metric on the pseudo-origin is $\alpha=\sqrt{2}\paren{\frac{1}{\pi}-2\lambda}$.
\end{prop}

Here, $\lambda$ again equals $\paren{\kappa/(n+1)^2}^{1/\paren{n-1}}$. Curiously, $\phi=\paren{\frac{1}{\pi}-2\lambda}\sqrt{2s_2}$ reduces the $\sigma$-measure compared to the Fubini--Study metric on the \emph{entire} Calabi--Yau, even though it was fit to produce flat metric only on a set of isolated points. We plot the performance of this formula as $\paren{\cdots}\sqrt{s_2}$ in \cref{fig:pareto_ansatz}.

\section{Discussion}

This paper seeks to shed light on the underlying analytic form for the K\"ahler potential of a Ricci-flat metric over a Calabi--Yau manifold. This is typically considered to be a \textquote{complicated non-holomorphic function} \cite{ashmore_machine_2020}. 

Our main contributions are as follows. We develop a novel approach for analysing numerical flat metric approximations, with a specific focus on Fermat manifolds due to their highly symmetric nature. We show that saliencies of neural network approximations yield new insights into symmetries of the flat metric of these manifolds. We pursue the theoretical aspect of this result and formulate a criterion for when the flat metric admits a representation in terms of fewer parameters than previously thought. The criterion concerns the symmetry group of $\phi$ when extended to the ambient space, and in particular says that if the only monomial in $Q$ containing some $Z_i$ is $Z_i^{n+1}$, then $\phi$ depends only on $\abs{Z_i}$ and not on $\arg Z_i$. We also explore practical consequences of this, and use our criterion to greatly enhance neural network performance by reducing training time and memory cost, while increasing accuracy.

Specifically, we demonstrate that isometries of Fermat Calabi--Yaus can always be encoded into NN models using Graph Neural Networks and a carefully chosen set of features that are invariant to these isometries. However, we also show that our criterion allows us to create ModNet, an even simpler model that converges to the same loss on Fermat manifolds after training for just one epoch. This calls for a deeper investigation into the criterion (\cref{thm:hidden_sym}), since it has substantial experimental support despite being left unproved. If this conjecture can be falsified, it would imply that some features have a large impact on the flat metric, while others have a proportionally very small yet non-zero impact. This hierarchy problem would then require a theoretical explanation, and would have practical consequences for approximating flat metrics. This is evidenced by our model, ModNet, which outperforms more general neural networks while using only a limited set of features.

In the second half of the paper we treat the flat metrics with symbolic methods. Through a combination of symbolic regression, least-squares fitting, and our own intuition, we distill our neural nets into short closed-form approximations with no increase in loss. These expressions have up to 2000x less parameters than the models they were distilled from (\cref{fig:pareto_ansatz}), and orders of magnitude less than the Hermitian matrix in the Donaldson's algorithm \cite{donaldson_numerical_2005}.

Crucially, our results are \emph{generalisable}. All of our conjectures (\cref{sec:hidden_sym}), observations (\cref{sec:phi_conjecture}), symbolic constructions (\cref{sec:phi_symb,sec:exact_formulae}), etc., hold equally well on every Calabi--Yau in the Fermat family that we tested, from the complex torus to the sextic. The consistency of our results rules out the chance that they are one-off coincidences. This motivates further study of these manifolds as a unified family with shared properties, rather than taking the more common approach of only looking at deformations of a Calabi--Yau in a single dimension. Furthermore, even though we give formulae for Fermat Calabi--Yaus only, they can be constructed for other symmetric hypersurfaces too, as their only precursor is using \cref{thm:hidden_sym} to reduce the search space. 

Finally, we use our high-quality approximations to propose a variety of novel statements about the flat metrics. We show that our flat metric approximation is very close to satisfying $c_3\leq0$ where $c_3$ is the third Chern form, or Euler density. We also show that our proposed constraints for the flat metric are sufficient to yield an exact solution on several loci on Fermat manifolds. In this way we again demonstrate the strength of our symmetry arguments.

In conclusion, our results are relevant both to algebraic geometry, by showing how Explainable AI produces new insights into properties of Calabi--Yau flat metrics, and to string phenomenology, by showing that it is possible for the flat metrics to be both highly accurate and interpretable. This paper barely scratches the surface of the consequences of our symmetry arguments, and we hope to inspire future work that will enable a better understanding of the flat metrics and the symmetries thereof.

\acknowledgments

We wish to thank Daattavya Aggrawal, Per Berglund, Carl Henrik Ek, Mario Garcia-Fernandez, Tristan H\"ubsch, Ferenc Husz\'ar, Vishnu Jejjala,  Dami\'an Mayorga Pe\~na, and Pietro Li\`o for helpful conversations. We also thank Justin Tan and Oisin Kim for their valuable comments and assistance with the benchmarking process. Part of this work was supported by a Turing AI World-Leading Researcher Fellowship G111021. Viktor Mirjani\'c is supported by a Trinity College studentship. Challenger Mishra is supported by the Accelerate Program for Scientific Discovery. 

\appendix
\section{Proofs}
\label{sec:proofs}

\subsection{Proof of Theorem \ref{thm:closed_form_w}}

\begin{lemma} An invertible matrix $A$ and vectors $u,v$ satisfy
        \begin{align*}
            \det\paren*{A+uv^\dag} &= \paren*{1+v^\dag A^{-1} u}\det \paren{A} && \text{(Matrix Determinant Lemma)}\\
            \paren*{A+uv^\dag}^{-1} &= A^{-1} - \frac{A^{-1}uv^\dag A^{-1}}{1+v^\dag A^{-1} u} && \text{(Sherman-Morrison Formula)}
        \end{align*}
\end{lemma}
\begin{lemma}[Euler's Homogeneous Function Theorem] An $n$-th degree homogeneous function $u$ satisfies
    \begin{equation*}
        \sum_i Z_i \frac{\partial u}{\partial Z_i} = n u
    \end{equation*}
\end{lemma}
\begin{lemma}[Pullback of Fubini Study Identity]
\label{lemm:pullbacked_fs_det}
    Suppose $u$ is a $(0,0)$-form of degree $n+1$ on $\mathbb{P}^n$ and let $\iota^*$ be the pullback to the variety defined by $u=0$. Then, on a patch where $Z_0=1$, and where we eliminate $Z_n$ via $u$ to obtain local coordinates $z$, we have 
    \begin{equation*}
        \det\iota^* g\FS=\frac{\norm{\nabla u}^2}{\abs*{\frac{\partial u}{\partial Z_n}}^2\paren*{1+\norm{z}^2}^n}
    \end{equation*}
    \begin{proof}
        First, define $u_i=\partial u/\partial Z_i$. Then, define the pullback matrix $J$ of $\iota^*$ as
        \begin{equation*}
            \renewcommand*{\arraystretch}{1.5}
            J = \begin{pmatrix}
                1 & \dots & 0 & -u_1 / u_n \\
                \vdots & \ddots & \vdots & \vdots \\
                0 & \dots & 1 & -u_{n-1} / u_n \\
            \end{pmatrix}_{(n-1)\times n}
        \end{equation*}
        and write $\iota^*g\FS = J g\FS J^\dag$. The main difficulty is in that $J$ is not square, so $\det\iota^*g\FS \neq \abs{\det J}^2 \det g\FS$. Instead, we use the matrix determinant lemma and Sherman-Morrison to obtain  
        \begin{equation*}
            \det\paren*{J J^\dag}= 
            \det\paren*{I+\frac{u_{1:n-1}u_{1:n-1}^\dag}{\abs*{u_n}^2}}= 
            1 + \frac{\sum_{i=1}^{n-1} \abs*{u_i}^2}{\abs*{u_n}^2}= 
            \frac{\norm*{u_{1:n}}^2}{\abs*{u_n}^2}
        \end{equation*}
        \begin{equation*}
            J^\dag \paren*{J J^\dag}^{-1} J= 
            J^\dag\paren*{I+\frac{u_{1:n-1}u_{1:n-1}^\dag}{\abs*{u_n}^2+u_{1:n-1}^\dag u_{1:n-1}}} J= 
            I-\frac{u_{1:n}u_{1:n}^\dag}{\norm*{u_{1:n}}^2}
        \end{equation*}
        where we use $u_{a:b}$ to specify a range of values in $u$. Finally, we compute
        {\allowdisplaybreaks
        \begin{align*}
            \det \iota^* g\FS &= \det \paren*{J g\FS J^\dag} && \text{substitute $\iota^*$}\\
            &=\det\paren*{J\paren*{\frac{1}{1+\norm{z}^2}I - \frac{z^\dag \otimes z}{\paren*{1+\norm{z}^2}^2}}J^\dag} && \text{substitute $g\FS$}\\
            &=\frac{1}{\paren*{1+\norm{z}^2}^{n-1}}\det\paren*{JJ^\dag - \frac{(J z^\dag) (z J^\dag)}{1+\norm{z}^2}} && \text{rearrange}\\
            &=\frac{1}{\paren*{1+\norm{z}^2}^{n-1}}\paren*{1 - \frac{z J^\dag \paren*{J J^\dag}^{-1} J z^\dag}{1+\norm{z}^2}}\det\paren*{JJ^\dag} && \text{matrix det lemma}\\
            &=\frac{\norm*{u_{1:n}}^2}{\abs*{u_n}^2\paren*{1+\norm{z}^2}^{n-1}} \paren*{1 - \frac{z J^\dag \paren*{J J^\dag}^{-1} J z^\dag}{1+\norm{z}^2}} && \text{substitute $\det\paren*{J J^\dag}$}\\
            &=\frac{\norm*{u_{1:n}}^2}{\abs*{u_n}^2\paren*{1+\norm{z}^2}^{n-1}} \paren*{1 - \frac{z \paren*{\norm*{u_{1:n}}^2 I - u_{1:n}u_{1:n}^\dag}z^\dag}{\norm*{u_{1:n}}^2\paren*{1+\norm{z}^2}}} && \text{substitute $J^\dag \paren*{J J^\dag}^{-1} J$}\\
            &=\frac{\norm*{u_{1:n}}^2}{\abs*{u_n}^2\paren*{1+\norm{z}^2}^{n-1}} \frac{\norm*{u_{1:n}}^2 + \abs*{z_{1:n} u_{1:n}}^2}{\norm*{u_{1:n}}^2\paren*{1+\norm{z}^2}} && \text{simplify $\paren{-}$}\\
            &=\frac{\norm*{u_{1:n}}^2 + \abs*{z_{1:n} u_{1:n}}^2}{\abs*{u_n}^2\paren*{1+\norm{z}^2}^{n}} && \text{factor}\\
            &=\frac{\norm*{u_{1:n}}^2 + \abs*{(n+1) u - u_0}^2}{\abs*{u_n}^2\paren*{1+\norm{z}^2}^{n-1}} && \text{Euler}\\
            &=\frac{\norm*{u_{1:n}}^2 + \abs*{u_0}^2}{\abs*{u_n}^2\paren*{1+\norm{z}^2}^{n}}&& \text{substitute $u=0$}\\
            &=\frac{\norm{\nabla u}^2}{\abs*{\frac{\partial u}{\partial Z_n}}^2\paren*{1+\norm{z}^2}^{n}} && \text{introduce $\nabla u$}
        \end{align*}
        }
        which completes the proof.
    \end{proof}
\end{lemma}
\begin{theorem}[Integration Weights Identity]
    Consider a Calabi--Yau hypersurface embedded into $\PP^n$ with the defining polynomial $Q$. Then, the integration weights $w$ satisfy
    \begin{equation*}
        w = \frac{\norm{Z}^{2n}}{\norm{\nabla Q}^2}
    \end{equation*}
    \begin{proof}
        From \cref{lemm:pullbacked_fs_det} we have 
        \begin{equation*}
            \det\iota^* g\FS=\frac{\norm{\nabla Q}^2}{\abs*{\frac{\partial Q}{\partial Z_n}}^2\paren*{1+\norm{z}^2}^n}
        \end{equation*}
        since $Q$ is homogeneous. We also know that
        \begin{equation*}
            \det g\fl = \frac{1}{\abs*{\frac{\partial Q}{\partial Z_n}}^2}
        \end{equation*}
        Taking the fraction and setting $\norm{Z}^2=1+\norm{z}^2$ completes the proof.
    \end{proof}
\end{theorem}

\subsection{Proof of Proposition \ref{prop:not_norms_in_general}}

\begin{prop}
    Consider a $d$-dimensional Fermat Calabi--Yau and an arbitrary $0$-form $f:\mathrm{CY}\to\RR$ with the symmetry group $\Sym(f)=\Isom(\mathrm{CY})$. Then, every such $f$ can be written as $f(Z,\overline{Z})=g(\abs{Z_1},\dots,\abs{Z_n})$ for some $g:\RR^{d+2}\to\RR$ if and only if $d=1$.
    \begin{proof}
        We will proceed by proving the statement for $d=1$ and providing a counterexample for $d>1$.
        \begin{itemize}
            \item Case $d=1$:
            
            The $1$-dimensional Fermat Calabi--Yau is the complex torus with $Q=1+z_1^3+z_2^3=0$, and $\Isom(\mathrm{CY})=\ZZ_2\times\ZZ_3^2\times\SS_3$. From now on we will ignore the permutation group $\SS_3$ and focus only on toric $\ZZ_3^2$ and conjugation $\ZZ_2$ symmetries.

            Firstly, by \cref{prop:f_rot_sym} $f$ must admit parameterisation in terms of $z_i^3$ due to toric symmetries of $Q$ and $f$, that is, due to invariance under rotation of $z_i$ by the $3^\text{rd}$ root of unity $\exp \frac{2i\pi}{3}$.
        
            Secondly, $\Re z_i^3$ are expressible in terms of norms via $Q=0$, for example:
            \begin{equation*}
                2 \Re z_1^3 = \abs{z_1^3+1}^2-\abs{z_1^3}^2-1 = \abs{Q-z_2^3}^2-\abs{z_1^3}^2-1 = \abs{z_2}^6 - \abs{z_1}^6 - 1
            \end{equation*}
            Thirdly, $\Im z_i^3=\pm \sqrt{\abs{z_i}^6 - \left(\Re z_i^3\right)^2}$ are also functions of norms, up to a $\pm$ sign, which is free and corresponds to the conjugation symmetry of $Q$ and $f$.
    
            Taking the above together, we obtain that $f$ is a function of the norms $\abs{z_i}$ as desired.

            \item Case $d>1$:

                    Suppose the homogeneous coordinates are $Z_1$ to $Z_n$, and consider the locus given by $Z_1(\theta)=e^{i\theta}$, $Z_2(\theta)=e^{i\theta+\frac{i\pi}{n}}$, $Z_{k>2} = e^{\frac{2(k-2)i\pi}{(n-2)^2}}$ for $\theta\in[0,2\pi)$. One can see that $\sum Z_k^n=0$ by summing the roots of unity, so this locus indeed lies on the Fermat manifold. 
                    
                    The locus satisfies $\abs{Z_k}=1$ for all $k$. Therefore, any function that depends only on the absolute values would be constant there.

                    Now define $f:\mathrm{CY}\to\RR$ as $f(Z,\overline{Z})=\cos\arg Z_1^n$. It clearly has both the toric $\ZZ_n^{n-1}$ and conjugation $\ZZ_2$ symmetries. On the locus it simplifies to $\cos n\theta$ which is clearly not constant. Therefore, $f$ cannot be written in terms of absolute values only.
        \end{itemize}
    \end{proof}
\end{prop}

\section{Low Loss Coefficients}
\label{sec:coeffs}

In \cref{sec:phi_symb} we introduced the formula
\begin{equation*}
    \hat\phi\approx \sum_i\sum_j c_{ij}\sqrt[n+1]{s_i}^j
\end{equation*}
that not only approximates $\hat\phi$ well, but simultaneously achieves low $\sigma$-loss. Here we present exact coefficients computed with the least squares method.

Observe how the coefficients are spanning several orders of magnitude. Also, note that the results are quite sensitive to the precision of $c_{ij}$, and that truncating them increases loss.
\begin{table}[htbp]
    \centering
    \caption{Coefficients $c_{ij}$ for the torus ($d=1$), achieving $\mathcal{L}_\sigma=\num{3.6e-5}$}
    \begin{tabular}{cccc}
    \toprule
    \multicolumn{1}{c}{} & \multicolumn{3}{c}{$j$} \\
    \cmidrule(rl){2-4}
         { $i$ } & {$1$} & {$2$} & {$3$} \\
         
    \cmidrule(r){1-1} \cmidrule(rl){2-4}
    2 & \num{0.00151509} & \num{0.10174207} & \num{-0.02850138} \\
    3 & \num{0.00029161} & \num{0.01253320} & \num{-0.00450205} \\
    4 & \num{-0.00069893} & \num{-0.01621588} & \num{0.00229031} \\
    \bottomrule
    \end{tabular}
\end{table}
 
\begin{table}[htbp]
    \centering
    \caption{Coefficients $c_{ij}$ for the quintic ($d=3$), achieving $\mathcal{L}_\sigma=\num{0.0011}$}
    \begin{tabular}{ccccc}
    \toprule
    \multicolumn{1}{c}{} & \multicolumn{4}{c}{$j$} \\
    \cmidrule(rl){2-5}
         { $i$ } & {$1$} & {$2$} & {$3$} & {$4$} \\
         
    \cmidrule(r){1-1} \cmidrule(rl){2-5}
    2 & \num{4.79419e-01} & \num{-6.11943e-01} & \num{5.60522e-01} & \num{-1.84679e-01} \\
    3 & \num{6.50726e-02} & \num{2.39588e-02} & \num{1.79043e-04} & \num{-5.57522e-04} \\
    4 & \num{-7.71605e-05} & \num{4.15851e-02} & \num{3.7261e-03} & \num{-1.72934e-02} \\
    5 & \num{-2.46564e-02} & \num{-2.90067e-02} & \num{2.20076e-02} & \num{1.57947e-02} \\
    6 & \num{1.90515e-02} & \num{-7.08450e-03} & \num{3.45797e-05} & \num{-5.41703e-02} \\
    \bottomrule
    \end{tabular}
\end{table}
\section{Generating Interpretable Formulae with PySR}
\label{sec:pysr}

We attempted to use PySR by \textcite{cranmerInterpretableMachineLearning2023} in two different ways to find symbolic flat metrics.

Firstly, we tried to learn symbolic $\hat\phi$ in the standard knowledge distillation setup, where a pretrained MLP would act as the teacher, and PySR would be the student learning to approximate $\phi$. While PySR did find formulae that minimised the surrogate MSE loss on $\hat\phi$, these formulae did not minimise the underlying $\sigma$-loss on $\hat g$, so we abandoned this approach.\footnote{Do not forget that the $\sigma$-loss is itself a surrogate loss for the true Ricci loss, and alignment between those two is a problem of its own \cite{larfors_learning_2021,hendi_learning_2024}.}

Secondly, we attempted to learn symbolic $\hat\phi$ in an \textquote{end-to-end} fashion, by writing a custom Julia loss function that directly minimises the $\sigma$-loss, and therefore bypassing the student-teacher setup and the misalignment between approximations of $\phi$ and $g$.

This worked better, and we obtained several symbolic approximations on the torus (\cref{tab:symb_form_torus}). These formulae were found using a dataset of \textbf{just 300 points}, versus \num{e7} used in ML, demonstrating the strength of the simplicity bias that PySR offers. In particular, note that the last formula has loss comparable to the one found in \cref{sec:interpretable}.

\begin{table}[htbp]
    \centering
    \caption{Symbolic formulae for $\phi$ on the Complex Torus as discovered with PySR, and their associated $\sigma$-losses. The coefficients were slightly cleaned by hand. Note that $s_2$ alone is not enough to minimise $\sigma$-loss.}
    \begin{tabular}{cc}
    \toprule
         {Symbolic $\hat\phi$} & {$\mathcal{L}_\sigma$} \\
    \midrule
         $\num{0.0751} \times s_2$ & \num{0.0038} \\
         $\num{0.0804} \times s_2^\frac{3}{4}$ & \num{7.6449e-04} \\
         $\num{0.0961} \times s_2 \left(\frac{3}{4}\right)^{s_2}$& \num{7.5138e-04} \\
         $\left(\num{0.00501} + \num{0.01102}\times s_2 \num{-0.00157}\times s_3\right)^{\frac{1}{\pi}}$ & \num{4.4357e-05} \\
    \bottomrule
    \end{tabular}
    \label{tab:symb_form_torus}
\end{table}

However, this approach proved impossible to scale to higher dimensions due to differentiation bottlenecks in Julia code, so we could not test this for dimensions $d>1$. As for the formulae themselves, only the last one in \cref{tab:symb_form_torus} makes sense to generalise, and we observe that the generalisations work reasonably well.

\section{Feature Attribution on the Quintic}
\label{sec:quintic_attrib}

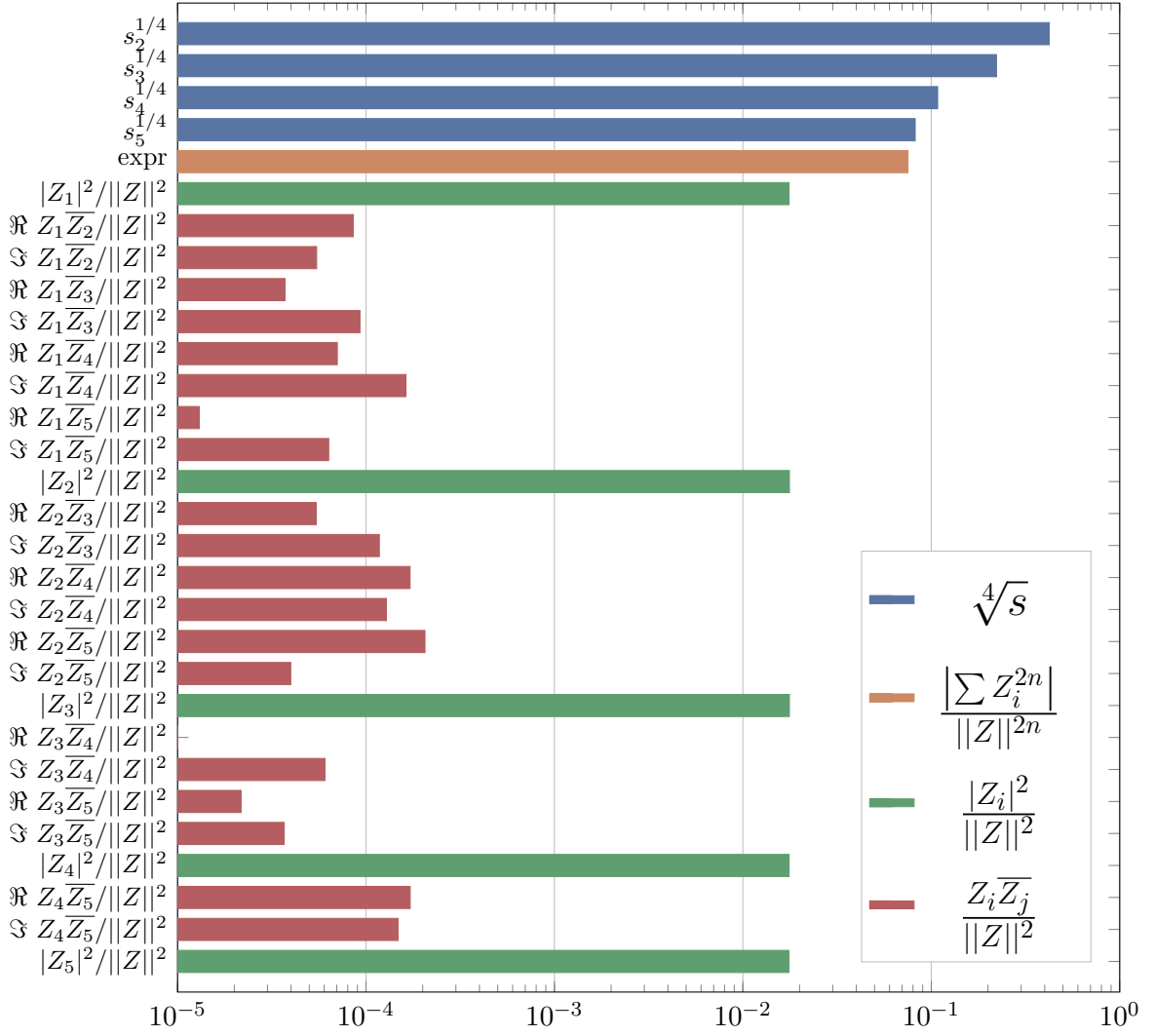
\begin{figure}[htbp]
    \centering
    \pgfplotstableread{
value label
4.2273355e-01 $s_2^{1/4}$
2.2192940e-01 $s_3^{1/4}$
1.0813779e-01 $s_4^{1/4}$
8.2155488e-02 $s_5^{1/4}$
}\attribblue

\pgfplotstableread{
value label
7.5228736e-02 expr
}\attriborange

\pgfplotstableread{
value label
1.7586568e-02 $|Z_1|^2/||Z||^2$
1.7667253e-02 $|Z_2|^2/||Z||^2$
1.7660448e-02 $|Z_3|^2/||Z||^2$
1.7602026e-02 $|Z_4|^2/||Z||^2$
1.7554171e-02 $|Z_5|^2/||Z||^2$
}\attribgreen

\pgfplotstableread{
value label
8.5540494e-05 $\Re\ Z_1\overline{Z_2}/||Z||^2$
3.7181428e-05 $\Re\ Z_1\overline{Z_3}/||Z||^2$
7.0424096e-05 $\Re\ Z_1\overline{Z_4}/||Z||^2$
1.3047736e-05 $\Re\ Z_1\overline{Z_5}/||Z||^2$
5.4450898e-05 $\Re\ Z_2\overline{Z_3}/||Z||^2$
1.7136396e-04 $\Re\ Z_2\overline{Z_4}/||Z||^2$
2.0553534e-04 $\Re\ Z_2\overline{Z_5}/||Z||^2$
8.7263461e-06 $\Re\ Z_3\overline{Z_4}/||Z||^2$
2.1728085e-05 $\Re\ Z_3\overline{Z_5}/||Z||^2$
1.7157631e-04 $\Re\ Z_4\overline{Z_5}/||Z||^2$
5.4695272e-05 $\Im\ Z_1\overline{Z_2}/||Z||^2$
9.2899427e-05 $\Im\ Z_1\overline{Z_3}/||Z||^2$
1.6299685e-04 $\Im\ Z_1\overline{Z_4}/||Z||^2$
6.3392290e-05 $\Im\ Z_1\overline{Z_5}/||Z||^2$
1.1768050e-04 $\Im\ Z_2\overline{Z_3}/||Z||^2$
1.2832067e-04 $\Im\ Z_2\overline{Z_4}/||Z||^2$
3.9860155e-05 $\Im\ Z_2\overline{Z_5}/||Z||^2$
6.0567130e-05 $\Im\ Z_3\overline{Z_4}/||Z||^2$
3.6799138e-05 $\Im\ Z_3\overline{Z_5}/||Z||^2$
1.4786349e-04 $\Im\ Z_4\overline{Z_5}/||Z||^2$
}\attribred

\begin{tikzpicture}
    \pgfplotsset{    
        width=0.95\linewidth,
        height=15cm,
        every axis y label/.style={
            at={(current axis.above origin)},
            anchor=north east,
        },
        every y tick label/.append style={font=\small}
    }
    \begin{axis}[
        log origin=infty,
        xmode=log, 
        xmin=10^-5, xmax=1,
        xmajorgrids=true,
        enlarge y limits=0.033,
        y dir=reverse,
        legend pos=south east, 
        legend style={draw=gray!50, 
                        font=\LARGE, 
                        nodes={inner sep=9pt}},
        bar width=0.3cm,
        symbolic y coords={$s_2^{1/4}$,
                         $s_3^{1/4}$,
                         $s_4^{1/4}$,
                         $s_5^{1/4}$,
                         expr,
                         $|Z_1|^2/||Z||^2$,
                         $\Re\ Z_1\overline{Z_2}/||Z||^2$,
                         $\Im\ Z_1\overline{Z_2}/||Z||^2$,
                         $\Re\ Z_1\overline{Z_3}/||Z||^2$,
                         $\Im\ Z_1\overline{Z_3}/||Z||^2$,
                         $\Re\ Z_1\overline{Z_4}/||Z||^2$,
                         $\Im\ Z_1\overline{Z_4}/||Z||^2$,
                         $\Re\ Z_1\overline{Z_5}/||Z||^2$,
                         $\Im\ Z_1\overline{Z_5}/||Z||^2$,
                         $|Z_2|^2/||Z||^2$,
                         $\Re\ Z_2\overline{Z_3}/||Z||^2$,
                         $\Im\ Z_2\overline{Z_3}/||Z||^2$,
                         $\Re\ Z_2\overline{Z_4}/||Z||^2$,
                         $\Im\ Z_2\overline{Z_4}/||Z||^2$,
                         $\Re\ Z_2\overline{Z_5}/||Z||^2$,
                         $\Im\ Z_2\overline{Z_5}/||Z||^2$,
                         $|Z_3|^2/||Z||^2$,
                         $\Re\ Z_3\overline{Z_4}/||Z||^2$,
                         $\Im\ Z_3\overline{Z_4}/||Z||^2$,
                         $\Re\ Z_3\overline{Z_5}/||Z||^2$,
                         $\Im\ Z_3\overline{Z_5}/||Z||^2$,
                         $|Z_4|^2/||Z||^2$,
                         $\Re\ Z_4\overline{Z_5}/||Z||^2$,
                         $\Im\ Z_4\overline{Z_5}/||Z||^2$,
                         $|Z_5|^2/||Z||^2$,
                         },
        ytick={$s_2^{1/4}$,
                         $s_3^{1/4}$,
                         $s_4^{1/4}$,
                         $s_5^{1/4}$,
                         expr,
                         $|Z_1|^2/||Z||^2$,
                         $\Re\ Z_1\overline{Z_2}/||Z||^2$,
                         $\Im\ Z_1\overline{Z_2}/||Z||^2$,
                         $\Re\ Z_1\overline{Z_3}/||Z||^2$,
                         $\Im\ Z_1\overline{Z_3}/||Z||^2$,
                         $\Re\ Z_1\overline{Z_4}/||Z||^2$,
                         $\Im\ Z_1\overline{Z_4}/||Z||^2$,
                         $\Re\ Z_1\overline{Z_5}/||Z||^2$,
                         $\Im\ Z_1\overline{Z_5}/||Z||^2$,
                         $|Z_2|^2/||Z||^2$,
                         $\Re\ Z_2\overline{Z_3}/||Z||^2$,
                         $\Im\ Z_2\overline{Z_3}/||Z||^2$,
                         $\Re\ Z_2\overline{Z_4}/||Z||^2$,
                         $\Im\ Z_2\overline{Z_4}/||Z||^2$,
                         $\Re\ Z_2\overline{Z_5}/||Z||^2$,
                         $\Im\ Z_2\overline{Z_5}/||Z||^2$,
                         $|Z_3|^2/||Z||^2$,
                         $\Re\ Z_3\overline{Z_4}/||Z||^2$,
                         $\Im\ Z_3\overline{Z_4}/||Z||^2$,
                         $\Re\ Z_3\overline{Z_5}/||Z||^2$,
                         $\Im\ Z_3\overline{Z_5}/||Z||^2$,
                         $|Z_4|^2/||Z||^2$,
                         $\Re\ Z_4\overline{Z_5}/||Z||^2$,
                         $\Im\ Z_4\overline{Z_5}/||Z||^2$,
                         $|Z_5|^2/||Z||^2$,
                         },
    ]

    \addplot+[
        xbar,
        mark=none,
        color=deepBlue,
        fill=deepBlue,
    ] 
    table [y=label, x=value, col sep=comma] {\attribblue};
    
    \addplot+[
        xbar,
        mark=none,
        color=deepOrange,
        fill=deepOrange,
    ] 
    table [y=label, x=value, col sep=comma] {\attriborange};
    
    \addplot+[
        xbar,
        mark=none,
        color=deepGreen,
        fill=deepGreen,
    ] 
    table [y=label, x=value, col sep=comma] {\attribgreen};

    \addplot+[
        xbar,
        mark=none,
        color=deepRed,
        fill=deepRed,
    ] 
    table [y=label, x=value, col sep=comma] {\attribred};
    
    \legend{$\sqrt[4]{s}$, 
    $\frac{ \left|\sum Z_i^{2n}\right|}{||Z||^{2n}}$, 
    $\frac{|Z_i|^2}{||Z||^2}$,
    $\frac{Z_i \overline{Z_j}}{||Z||^2}$} 

    \end{axis}
\end{tikzpicture}
    \caption{Normalised attribution weights $a$ on the quintic displayed on the log scale, computed by averaging over \num{2e5} points.}
    \label{fig:quintic_attrib}
\end{figure}

See \cref{fig:quintic_attrib}.

\section{Second Order Differentiation Lemma}
\label{sec:chain_rule}

\begin{lemma}[Chain Rule for Second-Order Mixed Wirtinger Derivatives]
Let $f: \mathbb{C}^n \to \mathbb{R}^m$ and $g: \mathbb{R}^m \to \mathbb{R}$ be functions such that
\begin{align*}
    z &= (z_1, \ldots, z_n) \in \mathbb{C}^n \\
    f(z, \overline{z}) &= (f_1, \ldots, f_m) \\
    h(z, \overline{z}) &= g(f(z, \overline{z}))
\end{align*}
Then, for any $j,k = 1, \ldots, n$, the second-order mixed Wirtinger derivative of $h$ is given by:
\begin{equation}
    \frac{\partial^2 h}{\partial z_j \partial \overline{z_k}} = \sum_{i=1}^m \sum_{l=1}^m \frac{\partial^2 g}{\partial f_i \partial f_l} \frac{\partial f_i}{\partial z_j} \frac{\partial f_l}{\partial \overline{z_k}} + \sum_{i=1}^m \frac{\partial g}{\partial f_i} \frac{\partial^2 f_i}{\partial z_j \partial \overline{z_k}}
\end{equation}
\end{lemma}



\printbibliography

@article{constantin2016family,
  title={The family problem: hints from heterotic line bundle models},
  author={Constantin, Andrei and Lukas, Andre and Mishra, Challenger},
  journal={Journal of High Energy Physics},
  volume={2016},
  number={3},
  pages={1--17},
  year={2016},
  publisher={Springer}
}

@article{candelas2016hodge,
  title={Hodge numbers for CICYs with symmetries of order divisible by 4},
  author={Candelas, Philip and Constantin, Andrei and Mishra, Challenger},
  journal={Fortschritte der Physik},
  volume={64},
  number={6-7},
  pages={463--509},
  year={2016},
  publisher={Wiley Online Library}
}

@article{ibanez1992discrete,
  title={Discrete gauge symmetries and the origin of baryon and lepton number conservation in supersymmetric versions of the standard model},
  author={Ibanez, Luis E and Ross, Graham G},
  journal={Nuclear Physics B},
  volume={368},
  number={1},
  pages={3--37},
  year={1992},
  publisher={Elsevier}
}

@misc{bronstein_geometric_2021,
	title = {Geometric {Deep} {Learning}: {Grids}, {Groups}, {Graphs}, {Geodesics}, and {Gauges}},
	shorttitle = {Geometric {Deep} {Learning}},
	url = {http://arxiv.org/abs/2104.13478},
	urldate = {2023-05-17},
	publisher = {arXiv},
	author = {Bronstein, Michael M. and Bruna, Joan and Cohen, Taco and Veličković, Petar},
	month = may,
	year = {2021},
	note = {arXiv:2104.13478 [cs, stat]},
}

@article{cybenko_approximation_1989,
	title = {Approximation by superpositions of a sigmoidal function},
	volume = {2},
	issn = {1435-568X},
	url = {https://doi.org/10.1007/BF02551274},
	doi = {10.1007/BF02551274},
	language = {en},
	number = {4},
	urldate = {2023-04-10},
	journal = {Mathematics of Control, Signals and Systems},
	author = {Cybenko, G.},
	month = dec,
	year = {1989},
	pages = {303--314},
}

@article{frucht_herstellung_1938,
	title = {Herstellung von {Graphen} mit vorgegebener abstrakter {Gruppe}},
	volume = {6},
	issn = {0010-437X},
	language = {German},
	journal = {Compositio Mathematica},
	author = {Frucht, R.},
	year = {1938},
	pages = {239--250},
}

@article{jejjala_neural_2021,
	title = {Neural network approximations for {Calabi}-{Yau} metrics},
	volume = {2022},
	issn = {1029-8479},
	url = {https://doi.org/10.1007/JHEP08(2022)105},
	doi = {10.1007/JHEP08(2022)105},
	language = {en},
	number = {8},
	urldate = {2024-12-02},
	journal = {Journal of High Energy Physics},
	author = {Jejjala, Vishnu and Peña, Damián Kaloni Mayorga and Mishra, Challenger},
	month = aug,
	year = {2022},
	pages = {105},
}

@article{douglas_numerical_2008,
	title = {Numerical {Calabi}-{Yau} metrics},
	volume = {49},
	issn = {0022-2488, 1089-7658},
	url = {http://arxiv.org/abs/hep-th/0612075},
	doi = {10.1063/1.2888403},
	number = {3},
	urldate = {2024-01-31},
	journal = {Journal of Mathematical Physics},
	author = {Douglas, Michael R. and Karp, Robert L. and Lukic, Sergio and Reinbacher, Rene},
	month = mar,
	year = {2008},
	note = {arXiv:hep-th/0612075},
	pages = {032302},
}

@misc{headrick_energy_2010,
	title = {Energy functionals for {Calabi}-{Yau} metrics},
	url = {http://arxiv.org/abs/0908.2635},
	urldate = {2024-01-31},
	publisher = {arXiv},
	author = {Headrick, Matthew and Nassar, Ali},
	month = feb,
	year = {2010},
	note = {arXiv:0908.2635 [hep-th]},
}

@misc{larfors_learning_2021,
	title = {Learning {Size} and {Shape} of {Calabi}-{Yau} {Spaces}},
	url = {http://arxiv.org/abs/2111.01436},
	urldate = {2023-02-21},
	publisher = {arXiv},
	author = {Larfors, Magdalena and Lukas, Andre and Ruehle, Fabian and Schneider, Robin},
	month = nov,
	year = {2021},
	note = {arXiv:2111.01436 [hep-th]},
}

@misc{berglund_machine_2023,
	title = {Machine {Learned} {Calabi}-{Yau} {Metrics} and {Curvature}},
	url = {http://arxiv.org/abs/2211.09801},
	doi = {10.48550/arXiv.2211.09801},
	urldate = {2023-03-04},
	publisher = {arXiv},
	author = {Berglund, Per and Butbaia, Giorgi and Hübsch, Tristan and Jejjala, Vishnu and Peña, Damián Mayorga and Mishra, Challenger and Tan, Justin},
	month = feb,
	year = {2023},
	note = {arXiv:2211.09801 [hep-th]},
}

@article{headrick_numerical_2005,
	title = {Numerical {Ricci}-flat metrics on {K3}},
	volume = {22},
	issn = {0264-9381, 1361-6382},
	url = {https://iopscience.iop.org/article/10.1088/0264-9381/22/23/002},
	doi = {10.1088/0264-9381/22/23/002},
	language = {en},
	number = {23},
	urldate = {2024-05-09},
	journal = {Classical and Quantum Gravity},
	author = {Headrick, Matthew and Wiseman, Toby},
	month = dec,
	year = {2005},
	pages = {4931--4960},
}

@misc{hendi_learning_2024,
	title = {Learning {Group} {Invariant} {Calabi}-{Yau} {Metrics} by {Fundamental} {Domain} {Projections}},
	url = {http://arxiv.org/abs/2407.06914},
	language = {en},
	urldate = {2024-07-23},
	publisher = {arXiv},
	author = {Hendi, Yacoub and Larfors, Magdalena and Walden, Moritz},
	month = jul,
	year = {2024},
	note = {arXiv:2407.06914 [hep-th, physics:math-ph]},
}

@article{cranmerDiscovering2020,
    title={Discovering Symbolic Models from Deep Learning with Inductive Biases},
    author={Miles Cranmer and Alvaro Sanchez-Gonzalez and Peter Battaglia and Rui Xu and Kyle Cranmer and David Spergel and Shirley Ho},
    journal={NeurIPS 2020},
    year={2020},
    eprint={2006.11287},
    archivePrefix={arXiv},
    primaryClass={cs.LG}
}

@misc{cranmerInterpretableMachineLearning2023,
    title = {Interpretable {Machine} {Learning} for {Science} with {PySR} and {SymbolicRegression}.jl},
    url = {http://arxiv.org/abs/2305.01582},
    doi = {10.48550/arXiv.2305.01582},
    urldate = {2023-07-17},
    publisher = {arXiv},
    author = {Cranmer, Miles},
    month = may,
    year = {2023},
    note = {arXiv:2305.01582 [astro-ph, physics:physics]},
}

@misc{kachru_k3_2020,
	title = {K3 metrics},
	url = {http://arxiv.org/abs/2006.02435},
	doi = {10.48550/arXiv.2006.02435},
	urldate = {2023-03-14},
	publisher = {arXiv},
	author = {Kachru, Shamit and Tripathy, Arnav and Zimet, Max},
	month = oct,
	year = {2020},
	note = {arXiv:2006.02435 [hep-th]},
}

@article{davies_advancing_2021,
	title = {Advancing mathematics by guiding human intuition with {AI}},
	volume = {600},
	copyright = {2021 The Author(s)},
	issn = {1476-4687},
	url = {https://www.nature.com/articles/s41586-021-04086-x},
	doi = {10.1038/s41586-021-04086-x},
	language = {en},
	number = {7887},
	urldate = {2023-03-04},
	journal = {Nature},
	author = {Davies, Alex and Veličković, Petar and Buesing, Lars and Blackwell, Sam and Zheng, Daniel and Tomašev, Nenad and Tanburn, Richard and Battaglia, Peter and Blundell, Charles and Juhász, András and Lackenby, Marc and Williamson, Geordie and Hassabis, Demis and Kohli, Pushmeet},
	month = dec,
	year = {2021},
	note = {Number: 7887
Publisher: Nature Publishing Group},
	pages = {70--74},
}

@article{ashmore_machine_2020,
	title = {Machine learning {Calabi}-{Yau} metrics},
	volume = {68},
	issn = {0015-8208, 1521-3978},
	url = {http://arxiv.org/abs/1910.08605},
	doi = {10.1002/prop.202000068},
	number = {9},
	urldate = {2023-02-15},
	journal = {Fortschritte der Physik},
	author = {Ashmore, Anthony and He, Yang-Hui and Ovrut, Burt},
	month = sep,
	year = {2020},
	note = {arXiv:1910.08605 [hep-th, stat]},
	pages = {2000068},
}

@article{donaldson_numerical_2005,
	title = {Some {Numerical} {Results} in {Complex} {Differential} {Geometry}},
	volume = {5},
    year = {2005},
	issn = {1558-8602},
	url = {https://link.intlpress.com/JDetail/1806164915712897025},
	doi = {10.4310/PAMQ.2009.v5.n2.a2},
	language = {EN},
	number = {2},
	urldate = {2024-12-02},
	journal = {Pure and Applied Mathematics Quarterly},
	author = {Donaldson, S. K.},
	note = {Publisher: International Press of Boston},
	pages = {571--618},
}

@misc{villar_scalars_2023,
	title = {Scalars are universal: {Equivariant} machine learning, structured like classical physics},
	shorttitle = {Scalars are universal},
	url = {http://arxiv.org/abs/2106.06610},
	language = {en},
	urldate = {2024-08-25},
	publisher = {arXiv},
	author = {Villar, Soledad and Hogg, David W. and Storey-Fisher, Kate and Yao, Weichi and Blum-Smith, Ben},
	month = feb,
	year = {2023},
	note = {arXiv:2106.06610 [math-ph, stat]},
}

@book{weyl_classical_1946,
	title = {The {Classical} {Groups}: {Their} {Invariants} and {Representations}},
	isbn = {978-0-691-05756-9},
	shorttitle = {The {Classical} {Groups}},
	language = {en},
	publisher = {Princeton University Press},
	author = {Weyl, Hermann},
	year = {1946},
	note = {Google-Books-ID: zmzKSP2xTtYC},
}

@misc{ek_calabi-yau_2024,
	title = {Calabi-{Yau} metrics through {Grassmannian} learning and {Donaldson}'s algorithm},
	url = {http://arxiv.org/abs/2410.11284},
	urldate = {2024-10-19},
	publisher = {arXiv},
	author = {Ek, Carl Henrik and Kim, Oisin and Mishra, Challenger},
	month = oct,
	year = {2024},
	note = {arXiv:2410.11284},
}

@inproceedings{sundararajan_axiomatic_2017,
	address = {Sydney, NSW, Australia},
	series = {{ICML}'17},
	title = {Axiomatic attribution for deep networks},
	urldate = {2024-12-02},
	booktitle = {Proceedings of the 34th {International} {Conference} on {Machine} {Learning} - {Volume} 70},
	publisher = {JMLR.org},
	author = {Sundararajan, Mukund and Taly, Ankur and Yan, Qiqi},
	month = aug,
	year = {2017},
	pages = {3319--3328},
}

@incollection{montavon_layer-wise_2019,
	address = {Cham},
	title = {Layer-{Wise} {Relevance} {Propagation}: {An} {Overview}},
	isbn = {978-3-030-28954-6},
	shorttitle = {Layer-{Wise} {Relevance} {Propagation}},
	url = {https://doi.org/10.1007/978-3-030-28954-6_10},
	language = {en},
	urldate = {2024-11-04},
	booktitle = {Explainable {AI}: {Interpreting}, {Explaining} and {Visualizing} {Deep} {Learning}},
	publisher = {Springer International Publishing},
	author = {Montavon, Grégoire and Binder, Alexander and Lapuschkin, Sebastian and Samek, Wojciech and Müller, Klaus-Robert},
	editor = {Samek, Wojciech and Montavon, Grégoire and Vedaldi, Andrea and Hansen, Lars Kai and Müller, Klaus-Robert},
	year = {2019},
	doi = {10.1007/978-3-030-28954-6_10},
	pages = {193--209},
}

@article{hornik_approximation_1991,
	title = {Approximation capabilities of multilayer feedforward networks},
	volume = {4},
	issn = {0893-6080},
	url = {https://doi.org/10.1016/0893-6080(91)90009-T},
	doi = {10.1016/0893-6080(91)90009-T},
	number = {2},
	urldate = {2024-11-13},
	journal = {Neural Netw.},
	author = {Hornik, Kurt},
	month = mar,
	year = {1991},
	pages = {251--257},
}

@misc{xu_how_2019,
	title = {How {Powerful} are {Graph} {Neural} {Networks}?},
	url = {http://arxiv.org/abs/1810.00826},
	language = {en},
	urldate = {2024-11-13},
	publisher = {arXiv},
	author = {Xu, Keyulu and Hu, Weihua and Leskovec, Jure and Jegelka, Stefanie},
	month = feb,
	year = {2019},
	note = {arXiv:1810.00826 [cs]},
}

@article{hornik_multilayer_1989,
	title = {Multilayer feedforward networks are universal approximators},
	volume = {2},
	issn = {0893-6080},
	url = {https://www.sciencedirect.com/science/article/pii/0893608089900208},
	doi = {10.1016/0893-6080(89)90020-8},
	number = {5},
	urldate = {2024-11-13},
	journal = {Neural Networks},
	author = {Hornik, Kurt and Stinchcombe, Maxwell and White, Halbert},
	month = jan,
	year = {1989},
	pages = {359--366},
}

@misc{raissi_physics_2017,
	title = {Physics {Informed} {Deep} {Learning} ({Part} {I}): {Data}-driven {Solutions} of {Nonlinear} {Partial} {Differential} {Equations}},
	shorttitle = {Physics {Informed} {Deep} {Learning} ({Part} {I})},
	url = {http://arxiv.org/abs/1711.10561},
	language = {en},
	urldate = {2024-11-13},
	publisher = {arXiv},
	author = {Raissi, Maziar and Perdikaris, Paris and Karniadakis, George Em},
	month = nov,
	year = {2017},
	note = {arXiv:1711.10561 [cs]},
}

@misc{berglund_cymyc_2024,
	title = {cymyc -- {Calabi}-{Yau} {Metrics}, {Yukawas}, and {Curvature}},
	url = {http://arxiv.org/abs/2410.19728},
	doi = {10.48550/arXiv.2410.19728},
	urldate = {2024-11-13},
	publisher = {arXiv},
	author = {Berglund, Per and Butbaia, Giorgi and Hübsch, Tristan and Jejjala, Vishnu and Mishra, Challenger and Peña, Damián Mayorga and Tan, Justin},
	month = oct,
	year = {2024},
	note = {arXiv:2410.19728},
}

@misc{liu_kan_2024,
	title = {{KAN}: {Kolmogorov}-{Arnold} {Networks}},
	shorttitle = {{KAN}},
	url = {http://arxiv.org/abs/2404.19756},
	doi = {10.48550/arXiv.2404.19756},
	urldate = {2024-11-15},
	publisher = {arXiv},
	author = {Liu, Ziming and Wang, Yixuan and Vaidya, Sachin and Ruehle, Fabian and Halverson, James and Soljačić, Marin and Hou, Thomas Y. and Tegmark, Max},
	month = jun,
	year = {2024},
	note = {arXiv:2404.19756},
}

@article{kolmogorov:superposition,
  author = {Kolmogorov, A. K.},
  journal = {Doklady Akademii Nauk SSSR},
  pages = {369--373},
  title = {On the Representation of Continuous Functions of Several Variables by Superposition of Continuous Functions of One Variable and Addition},
  volume = 114,
  year = 1957
}

@incollection{arnold_representation_2009,
	address = {Berlin, Heidelberg},
	title = {Representation of continuous functions of three variables by the superposition of continuous functions of two variables},
	isbn = {978-3-642-01742-1},
	url = {https://doi.org/10.1007/978-3-642-01742-1_6},
	language = {en},
	urldate = {2024-11-15},
	booktitle = {Collected {Works}: {Representations} of {Functions}, {Celestial} {Mechanics} and {KAM} {Theory}, 1957–1965},
	publisher = {Springer},
	editor = {Arnold, Vladimir I. and Givental, Alexander B. and Khesin, Boris A. and Marsden, Jerrold E. and Varchenko, Alexander N. and Vassiliev, Victor A. and Viro, Oleg Ya. and Zakalyukin, Vladimir M.},
	year = {2009},
	doi = {10.1007/978-3-642-01742-1_6},
	pages = {47--133},
}

@book{hubsch_calabi-yau_1994,
	title = {Calabi-{Yau} {Manifolds}: {A} {Bestiary} for {Physicists}},
	isbn = {978-981-02-1927-7},
	shorttitle = {Calabi-{Yau} {Manifolds}},
	language = {en},
	publisher = {World Scientific},
	author = {H\"ubsch, Tristan},
	year = {1994},
	note = {Google-Books-ID: bTRqDQAAQBAJ},
}

@misc{candelas_de_la_ossa,
	title = {Lectures on Complex Manifolds},
	url = {https://www.math.utoronto.ca/mgualt/courses/MAT477-2017/docs/Candelas-delaOssa.pdf},
	urldate = {2024-11-18},
	author = {Candelas, Philip and de la Ossa, Xenia},
	month = aug,
	year = {1987},
}

@misc{catanese_kummer_2021,
	title = {Kummer quartic surfaces, strict self-duality, and more},
	url = {http://arxiv.org/abs/2101.10501},
	doi = {10.48550/arXiv.2101.10501},
	language = {en},
	urldate = {2024-11-26},
	publisher = {arXiv},
	author = {Catanese, Fabrizio},
	month = may,
	year = {2021},
	note = {arXiv:2101.10501 [math]},
}

@incollection{gallier_isometries_2020,
	address = {Cham},
	title = {Isometries, {Local} {Isometries}, {Riemannian} {Coverings} and {Submersions}, and {Killing} {Vector} {Fields}},
	isbn = {978-3-030-46040-2},
	url = {https://doi.org/10.1007/978-3-030-46040-2_18},
	language = {en},
	urldate = {2024-11-26},
	booktitle = {Differential {Geometry} and {Lie} {Groups}: {A} {Computational} {Perspective}},
	publisher = {Springer International Publishing},
	author = {Gallier, Jean and Quaintance, Jocelyn},
	editor = {Gallier, Jean and Quaintance, Jocelyn},
	year = {2020},
	doi = {10.1007/978-3-030-46040-2_18},
	pages = {541--557},
}

@article{yau_ricci_1978,
	title = {On the ricci curvature of a compact kähler manifold and the complex monge-ampére equation, {I}},
	volume = {31},
	issn = {1097-0312},
	url = {https://onlinelibrary.wiley.com/doi/abs/10.1002/cpa.3160310304},
	doi = {10.1002/cpa.3160310304},
	language = {en},
	number = {3},
	urldate = {2023-03-04},
	journal = {Communications on Pure and Applied Mathematics},
	author = {Yau, Shing-Tung},
	year = {1978},
	pages = {339--411},
}

@article{candelas_vacuum_1985,
	title = {Vacuum configurations for superstrings},
	volume = {258},
	issn = {0550-3213},
	url = {https://www.sciencedirect.com/science/article/pii/0550321385906029},
	doi = {10.1016/0550-3213(85)90602-9},
	urldate = {2024-11-27},
	journal = {Nuclear Physics B},
	author = {Candelas, P. and Horowitz, Gary T. and Strominger, Andrew and Witten, Edward},
	month = jan,
	year = {1985},
	pages = {46--74},
}

@article{candelas_pair_1991,
	title = {A pair of {Calabi}-{Yau} manifolds as an exactly soluble superconformal theory},
	volume = {359},
	issn = {0550-3213},
	url = {https://www.sciencedirect.com/science/article/pii/0550321391902926},
	doi = {10.1016/0550-3213(91)90292-6},
	number = {1},
	urldate = {2024-11-28},
	journal = {Nuclear Physics B},
	author = {Candelas, Philip and De La Ossa, Xenia C. and Green, Paul S. and Parkes, Linda},
	month = jul,
	year = {1991},
	pages = {21--74},
}

@article{lukas_discrete_2020,
	title = {Discrete {Symmetries} of {Complete} {Intersection} {Calabi}–{Yau} {Manifolds}},
	volume = {379},
	issn = {1432-0916},
	url = {https://doi.org/10.1007/s00220-020-03838-6},
	doi = {10.1007/s00220-020-03838-6},
	language = {en},
	number = {3},
	urldate = {2024-12-02},
	journal = {Communications in Mathematical Physics},
	author = {Lukas, Andre and Mishra, Challenger},
	month = nov,
	year = {2020},
	pages = {847--865},
}

@article{candelas_highly_2018,
	title = {Highly {Symmetric} {Quintic} {Quotients}},
	volume = {66},
	copyright = {© 2018 WILEY-VCH Verlag GmbH \& Co. KGaA, Weinheim},
	issn = {1521-3978},
	url = {https://onlinelibrary.wiley.com/doi/abs/10.1002/prop.201800017},
	doi = {10.1002/prop.201800017},
	language = {en},
	number = {4},
	urldate = {2024-12-02},
	journal = {Fortschritte der Physik},
	author = {Candelas, Philip and Mishra, Challenger},
	year = {2018},
	note = {\_eprint: https://onlinelibrary.wiley.com/doi/pdf/10.1002/prop.201800017},
	pages = {1800017},
}

@article{candelas_new_2010,
	title = {New {Calabi}-{Yau} manifolds with small {Hodge} numbers},
	volume = {58},
	copyright = {Copyright © 2010 WILEY-VCH Verlag GmbH \& Co. KGaA, Weinheim},
	issn = {1521-3978},
	url = {https://onlinelibrary.wiley.com/doi/abs/10.1002/prop.200900105},
	doi = {10.1002/prop.200900105},
	language = {en},
	number = {4-5},
	urldate = {2024-11-28},
	journal = {Fortschritte der Physik},
	author = {Candelas, P. and Davies, R.},
	year = {2010},
	note = {\_eprint: https://onlinelibrary.wiley.com/doi/pdf/10.1002/prop.200900105},
	pages = {383--466},
}

@article{braun_free_2011,
	title = {On free quotients of complete intersection {Calabi}-{Yau} manifolds},
	volume = {2011},
	issn = {1029-8479},
	url = {https://doi.org/10.1007/JHEP04(2011)005},
	doi = {10.1007/JHEP04(2011)005},
	language = {en},
	number = {4},
	urldate = {2024-12-02},
	journal = {Journal of High Energy Physics},
	author = {Braun, Volker},
	month = apr,
	year = {2011},
	pages = {5},
}

@article{braun_calabi-yau_2008,
	title = {Calabi-{Yau} metrics for quotients and complete intersections},
	volume = {2008},
	issn = {1029-8479},
	url = {http://stacks.iop.org/1126-6708/2008/i=05/a=080?key=crossref.939d76450f934ae008a197ece9b0d8d1},
	doi = {10.1088/1126-6708/2008/05/080},
	language = {en},
	number = {05},
	urldate = {2024-12-02},
	journal = {Journal of High Energy Physics},
	author = {Braun, Volker and Brelidze, Tamaz and Douglas, Michael R and Ovrut, Burt A},
	month = may,
	year = {2008},
	pages = {080--080},
}

@phdthesis{davies_calabi-yau_2010,
	type = {{PhD} {Thesis}},
	title = {Calabi-{Yau} threefolds and heterotic string compactification},
	school = {Oxford U.},
	author = {Davies, Rhys},
	year = {2010},
}

@article{braun_three-generation_2010,
	title = {A three-generation {Calabi}-{Yau} manifold with small {Hodge} numbers},
	volume = {58},
	copyright = {Copyright © 2010 WILEY-VCH Verlag GmbH \& Co. KGaA, Weinheim},
	issn = {1521-3978},
	url = {https://onlinelibrary.wiley.com/doi/abs/10.1002/prop.200900106},
	doi = {10.1002/prop.200900106},
	language = {en},
	number = {4-5},
	urldate = {2024-12-02},
	journal = {Fortschritte der Physik},
	author = {Braun, V. and Candelas, P. and Davies, R.},
	year = {2010},
	note = {\_eprint: https://onlinelibrary.wiley.com/doi/pdf/10.1002/prop.200900106},
	pages = {467--502},
}

@article{candelas_calabi-yau_2018,
	title = {Calabi-{Yau} {Threefolds} with {Small} {Hodge} {Numbers}},
	volume = {66},
	copyright = {© 2018 WILEY-VCH Verlag GmbH \& Co. KGaA, Weinheim},
	issn = {1521-3978},
	url = {https://onlinelibrary.wiley.com/doi/abs/10.1002/prop.201800029},
	doi = {10.1002/prop.201800029},
	language = {en},
	number = {6},
	urldate = {2024-12-02},
	journal = {Fortschritte der Physik},
	author = {Candelas, Philip and Constantin, Andrei and Mishra, Challenger},
	year = {2018},
	note = {\_eprint: https://onlinelibrary.wiley.com/doi/pdf/10.1002/prop.201800029},
	pages = {1800029},
}

@article{raissi_physics-informed_2019,
	title = {Physics-informed neural networks: {A} deep learning framework for solving forward and inverse problems involving nonlinear partial differential equations},
	volume = {378},
	issn = {0021-9991},
	shorttitle = {Physics-informed neural networks},
	url = {https://www.sciencedirect.com/science/article/pii/S0021999118307125},
	doi = {10.1016/j.jcp.2018.10.045},
	urldate = {2024-12-02},
	journal = {Journal of Computational Physics},
	author = {Raissi, M. and Perdikaris, P. and Karniadakis, G. E.},
	month = feb,
	year = {2019},
	pages = {686--707},
}

@phdthesis{mishra_calabi-yau_2017,
	type = {{PhD} {Thesis}},
	title = {Calabi-{Yau} manifolds, discrete symmetries and string theory},
	url = {https://ora.ox.ac.uk/objects/uuid:4a174981-085e-4e81-8f27-b48533f08315},
	language = {English},
	urldate = {2024-12-02},
	school = {University of Oxford},
	author = {Mishra, C.},
	year = {2017},
}

@misc{bradbury_jax_2018,
	title = {{JAX}: composable transformations of {Python}+{NumPy} programs},
	url = {http://github.com/jax-ml/jax},
	author = {Bradbury, James and Frostig, Roy and Hawkins, Peter and Johnson, Matthew James and Leary, Chris and Maclaurin, Dougal and Necula, George and Paszke, Adam and VanderPlas, Jake and Wanderman-Milne, Skye and Zhang, Qiao},
	year = {2018},
}

@article{noauthor_ieee_2019,
	title = {{IEEE} {Standard} for {Floating}-{Point} {Arithmetic}},
	url = {https://ieeexplore.ieee.org/document/8766229},
	doi = {10.1109/IEEESTD.2019.8766229},
	urldate = {2024-12-05},
	journal = {IEEE Std 754-2019 (Revision of IEEE 754-2008)},
	month = jul,
	year = {2019},
	note = {Conference Name: IEEE Std 754-2019 (Revision of IEEE 754-2008)},
	pages = {1--84},
}

@article{douglas_numerical_2007,
	title = {Numerical solution to the hermitian {Yang}-{Mills} equation on the {Fermat} quintic},
	volume = {2007},
	issn = {1029-8479},
	url = {http://arxiv.org/abs/hep-th/0606261},
	doi = {10.1088/1126-6708/2007/12/083},
	language = {en},
	number = {12},
	urldate = {2024-08-12},
	journal = {Journal of High Energy Physics},
	author = {Douglas, Michael R. and Karp, Robert L. and Lukic, Sergio and Reinbacher, Rene},
	month = dec,
	year = {2007},
	note = {arXiv:hep-th/0606261},
	pages = {083--083},
}

@misc{gerdes_cyjax_2022,
	title = {{CYJAX}: {A} package for {Calabi}-{Yau} metrics with {JAX}},
	shorttitle = {{CYJAX}},
	url = {http://arxiv.org/abs/2211.12520},
	doi = {10.48550/arXiv.2211.12520},
	urldate = {2023-02-22},
	publisher = {arXiv},
	author = {Gerdes, Mathis and Krippendorf, Sven},
	month = nov,
	year = {2022},
	note = {arXiv:2211.12520 [hep-th]},
}

@misc{anderson_lectures_2023,
	title = {Lectures on {Numerical} and {Machine} {Learning} {Methods} for {Approximating} {Ricci}-flat {Calabi}-{Yau} {Metrics}},
	url = {http://arxiv.org/abs/2312.17125},
	doi = {10.48550/arXiv.2312.17125},
	urldate = {2024-12-07},
	publisher = {arXiv},
	author = {Anderson, Lara B. and Gray, James and Larfors, Magdalena},
	month = dec,
	year = {2023},
	note = {arXiv:2312.17125 [hep-th]},
}

@misc{butbaia_physical_2024,
	title = {Physical {Yukawa} {Couplings} in {Heterotic} {String} {Compactifications}},
	url = {http://arxiv.org/abs/2401.15078},
	doi = {10.48550/arXiv.2401.15078},
	urldate = {2024-12-07},
	publisher = {arXiv},
	author = {Butbaia, Giorgi and Peña, Damián Mayorga and Tan, Justin and Berglund, Per and Hübsch, Tristan and Jejjala, Vishnu and Mishra, Challenger},
	month = may,
	year = {2024},
	note = {arXiv:2401.15078 [hep-th]},
}

@misc{bouchard_lectures_2007,
	title = {Lectures on complex geometry, {Calabi}-{Yau} manifolds and toric geometry},
	url = {http://arxiv.org/abs/hep-th/0702063},
	doi = {10.48550/arXiv.hep-th/0702063},
	urldate = {2024-11-28},
	publisher = {arXiv},
	author = {Bouchard, Vincent},
	month = feb,
	year = {2007},
	note = {arXiv:hep-th/0702063},
}

@article{witten_symmetry_1985,
	title = {Symmetry breaking patterns in superstring models},
	volume = {258},
	issn = {0550-3213},
	url = {https://www.sciencedirect.com/science/article/pii/0550321385906030},
	doi = {10.1016/0550-3213(85)90603-0},
	urldate = {2024-12-11},
	journal = {Nuclear Physics B},
	author = {Witten, Edward},
	month = jan,
	year = {1985},
	pages = {75--100},
}

@misc{constantin_fermion_2024,
	title = {Fermion {Masses} and {Mixing} in {String}-{Inspired} {Models}},
	url = {http://arxiv.org/abs/2410.17704},
	doi = {10.48550/arXiv.2410.17704},
	language = {en},
	urldate = {2024-12-13},
	publisher = {arXiv},
	author = {Constantin, Andrei and Fraser-Taliente, Cristofero S. and Harvey, Thomas R. and Leung, Lucas T. Y. and Lukas, Andre},
	month = oct,
	year = {2024},
	note = {arXiv:2410.17704 [hep-th]},
}

@phdthesis{anderson_heterotic_2008,
	title = {Heterotic and {M}-theory compactifications for string phenomenology},
	language = {eng},
	school = {University of Oxford},
	author = {Anderson, Lara Briana},
	collaborator = {Candelas, P. and Lukas, Andre and University of Oxford Mathematical, Physical {and} Life Sciences Division},
	year = {2008},
	note = {Book Title: Heterotic and M-theory compactifications for string phenomenology},
}

@article{denef_distributions_2004,
	title = {Distributions of flux vacua},
	volume = {2004},
	issn = {1029-8479},
	url = {http://arxiv.org/abs/hep-th/0404116},
	doi = {10.1088/1126-6708/2004/05/072},
	number = {05},
	urldate = {2024-12-19},
	journal = {Journal of High Energy Physics},
	author = {Denef, Frederik and Douglas, Michael R.},
	month = may,
	year = {2004},
	note = {arXiv:hep-th/0404116},
	pages = {072--072},
}

@misc{douglas_numerical_2021,
	title = {Numerical {Calabi}-{Yau} metrics from holomorphic networks},
	url = {http://arxiv.org/abs/2012.04797},
	doi = {10.48550/arXiv.2012.04797},
	urldate = {2024-12-20},
	publisher = {arXiv},
	author = {Douglas, Michael R. and Lakshminarasimhan, Subramanian and Qi, Yidi},
	month = may,
	year = {2021},
	note = {arXiv:2012.04797 [hep-th]},
}

@article{anderson_moduli-dependent_2021,
	title = {Moduli-dependent {Calabi}-{Yau} and {SU}(3)-structure metrics from machine learning},
	volume = {2021},
	issn = {1029-8479},
	url = {https://doi.org/10.1007/JHEP05(2021)013},
	doi = {10.1007/JHEP05(2021)013},
	language = {en},
	number = {5},
	urldate = {2023-03-09},
	journal = {Journal of High Energy Physics},
	author = {Anderson, Lara B. and Gerdes, Mathis and Gray, James and Krippendorf, Sven and Raghuram, Nikhil and Ruehle, Fabian},
	month = may,
	year = {2021},
	pages = {13},
}

@article{donaldson_scalar_2001,
	title = {Scalar {Curvature} and {Projective} {Embeddings}, {I}},
	volume = {59},
	issn = {0022-040X},
	url = {https://projecteuclid.org/journals/journal-of-differential-geometry/volume-59/issue-3/Scalar-Curvature-and-Projective-Embeddings-I/10.4310/jdg/1090349449.full},
	doi = {10.4310/jdg/1090349449},
	number = {3},
	urldate = {2024-01-11},
	journal = {Journal of Differential Geometry},
	author = {Donaldson, S. K.},
	month = nov,
	year = {2001},
	note = {Publisher: Lehigh University},
	pages = {479--522},
}

\end{document}